\documentclass[10pt,aps,prd,fleqn,superscriptaddress,notitlepage,nofootinbib,preprintnumbers,nobalancelastpage]{revtex4-1}
\pdfoutput=1
\usepackage{amsmath,amssymb,graphicx,xspace,tikz}
\usepackage{placeins}
\newcommand{\eps}{\varepsilon}
\newcommand{\Alaric}{{\sc Alaric}\xspace}

\newcommand{\Sherpa}{{\sc Sherpa}\xspace}
\usepackage{hyperref}
\hypersetup{hidelinks,
  pdfauthor={Stefan Hoeche,Frank Krauss,Daniel Reichelt},
  pdftitle={The Alaric parton shower for hadron colliders},
  pdfkeywords={Parton Shower,QCD Resummation}
}

\newcommand{\affFermilab}{Fermi National Accelerator Laboratory, Batavia, IL, 60510, USA}
\newcommand{\affIPPP}{Institute for Particle Physics Phenomenology, Durham University, Durham DH1 3LE, UK}

\begin{document}
\preprint{FERMILAB-PUB-24-0178-T, IPPP/24/20, MCNET-24-07}
\title{The Alaric parton shower for hadron colliders}
\author{Stefan~H{\"o}che}\affiliation{\affFermilab}
\author{Frank~Krauss}\affiliation{\affIPPP}
\author{Daniel~Reichelt}\affiliation{\affIPPP}

\begin{abstract}
  We introduce the \Alaric parton shower for simulating QCD radiation
  at hadron colliders and present numerical results from an implementation
  in the event generator \Sherpa. \Alaric provides a consistent framework
  to quantify certain systematic uncertainties which cannot be eliminated
  by comparing the parton shower with analytic resummation. In particular, 
  it allows to study recoil effects away from the soft and collinear limits
  without the need to change the evolution variable or the splitting functions.
  We assess the performance of \Alaric in Drell-Yan lepton pair and QCD jet
  production, and present the first multi-jet merging for the new algorithm.
\end{abstract}

\maketitle

\section{Introduction}
Experiments at high-energy hadron colliders such as the CERN Large
Hadron Collider (LHC) have been the source of much of our
understanding of the smallest building blocks of matter.  While they
often do not reach the same precision as lepton colliders,
proton-(anti)proton machines offer unprecedented reach in available
center-of-mass energy, and thus open a pathway to the observation of
hitherto unknown particles as well as new interactions~\cite{
 EuropeanStrategyforParticlePhysicsPreparatoryGroup:2019qin,
 Narain:2022qud,Butler:2023glv}.
Quite naturally, opportunity comes at a cost.  The composite nature of
the beam particles, and the complex phenomenology of QCD at low and
high scales hinder the extraction of rare hadron-level signals from
large and often poorly understood backgrounds.  Computer simulations
in the form of Monte-Carlo event generators have so far proven the
only effective approach to this
problem~\cite{Buckley:2011ms,Campbell:2022qmc}.  Among the many
components of these event generators, the approximation of QCD
radiative corrections to all orders in perturbation theory is one of
the most important.  This component is implemented by parton showers.

The discovery of the gluon at {\sc Petra} about forty years ago
spurred the development of the first parton showers~\cite{Webber:1983if,
  Bengtsson:1986gz,Bengtsson:1986et,Marchesini:1987cf}.
Since then, the increasing center-of-mass energy of the experiments
mandated a corresponding increase in precision of the simulations,
which led to the development of spin correlation algorithms~\cite{
 Collins:1987cp,Knowles:1987cu,Knowles:1988vs,Knowles:1988hu},
matching to next-to-leading order fixed-order calculations~\cite{
 Frixione:2002ik,Nason:2004rx,Frixione:2007vw,Alioli:2010xd,
 Hoeche:2010pf,Hoeche:2011fd,Alwall:2014hca}
and the merging of calculations for varying jet
multiplicity~\cite{Catani:2001cc,Lonnblad:2001iq,Krauss:2002up,
  Alwall:2007fs,Hoeche:2009rj,Lonnblad:2011xx,Gehrmann:2012yg,
  Hoeche:2012yf,Frederix:2012ps,Lonnblad:2012ix,Platzer:2012bs}.
Color coherent parton evolution, manifesting itself through angular
ordering for global observables~\cite{Mueller:1981ex,Ermolaev:1981cm,
  Dokshitzer:1982fh,Dokshitzer:1982xr,Dokshitzer:1982ia,Bassetto:1982ma},
became a guiding principle for the construction of many early parton
shower algorithms~\cite{Webber:1986mc,Bengtsson:1986hr} and remains a
powerful computational tool. However, for observables sensitive to certain
correlations among partons and jets, angular ordering does not capture
all details of QCD radiative effects~\cite{Dasgupta:2001sh}.  This
class of observables can be better described by algorithms based on
the color dipole picture, first proposed and implemented
in~\cite{Gustafson:1987rq,Andersson:1989ki,Lonnblad:1992tz}, and later
extended to a more efficient and precise simulation framework~\cite{
  Giele:2011cb,Fischer:2016vfv,Brooks:2020upa,
  Brooks:2021kji,Campbell:2021svd}.
Algorithms based on the dipole picture were also supplemented by a
matching to single parton evolution in the collinear
limit~\cite{Nagy:2005aa,Nagy:2006kb,Schumann:2007mg,Giele:2007di,
  Platzer:2009jq,Hoche:2015sya,Cabouat:2017rzi}.
Most parton showers currently used by the LHC experiments are based on
this paradigm~\cite{Campbell:2022qmc}.  Recently, they have again been
revised, in order to achieve consistency with analytic resummation in
the limit of large center-of-mass energies~\cite{Dasgupta:2018nvj}.
The resulting improvements concern kinematic recoil effects~\cite{
  Dasgupta:2020fwr,Bewick:2019rbu,Forshaw:2020wrq,Nagy:2020rmk,
  Nagy:2020dvz,Dasgupta:2020fwr,vanBeekveld:2022zhl,Herren:2022jej,
  Assi:2023rbu,Preuss:2024vyu},
and an improved simulation of color coherence~\cite{Gustafson:1992uh,
  Giele:2011cb,Nagy:2012bt,Platzer:2012np,Nagy:2014mqa,Nagy:2015hwa,
  Platzer:2018pmd,Isaacson:2018zdi,Nagy:2019rwb,Nagy:2019pjp,
  Forshaw:2019ver,Hoche:2020pxj,DeAngelis:2020rvq,
  Holguin:2020joq,Hamilton:2020rcu}.

In this publication we will report on the extension of one of the new
dipole-like parton shower algorithms, called \Alaric~\cite{Herren:2022jej,
  Assi:2023rbu}, to initial-state radiation. A unique aspect of the
\Alaric method is the non-trivial dependence of splitting functions
on the azimuthal emission angle, even when spin correlations are not
included.  This allows to simulate the complete one-loop soft radiation
pattern without the need for angular ordering. At the same time, the
choice of recoil momentum necessary to implement four-momentum conservation
and on-shell conditions is left arbitrary, enabling an easy matching of
the parton-shower to analytic calculations for specific observables.
The new method satisfies the stringent criteria for next-to-leading
logarithmic (NLL) precision at leading color~\cite{Dasgupta:2018nvj}
for all recursively infrared safe observables~\cite{Herren:2022jej}.
Here we will discuss specifically the treatment of the collinear 
splitting functions in the context of different kinematics mappings,
focusing on terms which are not determined by the matching to a
soft eikonal. Sub-leading power corrections to these terms vanish
in the NLL limit, but can play a significant role at finite transverse
momentum~\cite{Hoeche:2017jsi} and must therefore be implemented
as faithful as possible. They are often important at hadron colliders
due to the enhanced gluon distribution at high energies and
small $x$~\cite{Jones:2017giv,ATLAS:2021yza}. 

The manuscript is structured as follows.
Section~\ref{sec:splitting_functions} introduces the collinear
splitting functions and presents a kinematics-independent definition
of the purely collinear terms for final- and initial-state evolution.
Section~\ref{sec:kinematics} introduces the kinematic mappings used in
our algorithm and discusses an extension of the proposal in
Ref.~\cite{Herren:2022jej}. Section~\ref{sec:results} presents some
first example phenomenological predictions in comparison to
experimental data from the Large Hadron Collider.  Finally,
Sec.~\ref{sec:outlook} discusses further directions of development.

\section{Splitting functions}
\label{sec:splitting_functions}
The precise form of the splitting functions is one of the main
systematic uncertainties in any parton-shower simulation.  Stringent
criteria exist only for the leading terms in gluon energy in the soft
gluon limit, and for the leading terms in transverse momentum in the
collinear limit.  These terms are determined by the known
soft~\cite{Bassetto:1984ik} and collinear~\cite{Dokshitzer:1977sg,
  Gribov:1972ri,Lipatov:1974qm,Altarelli:1977zs,Catani:2000ef,Catani:2002hc}
factorization properties of QCD amplitudes.  It is often assumed that
away from the limits, the splitting function can be used as is, without
the need to account for the precise definition of the splitting variable.
While it is certainly true that changes in its definition only induce
sub-leading corrections (of higher power in the soft or collinear
expansion parameter), the precise definition of the splitting kernels
plays an important role and can be used to capture non-leading effects.
A prominent example is the sub-leading power correction to the soft
splitting function~\cite{Low:1958sn,Burnett:1967km,DelDuca:1990gz},
which originates in classical radiative effects~\cite{Dokshitzer:2005bf}
and extends the naive soft limit to a physically more meaningful result.
Corrections of this type should clearly be included due to their
importance for the physics performance of the Monte-Carlo simulation.
A similarly important point is that the collinear splitting functions
can be computed as off-shell matrix elements in a
physical gauge~\cite{Catani:1999ss}, which implies that they contain
information on the structure of QCD amplitudes beyond the collinear
limit. If this structure is to be retained, it is necessary that the
splitting functions be evaluated with the exact same definition of
splitting variable that was used in their derivation. A change in the
kinematics parametrization must lead to identical physics predictions,
but it may require a different form of the splitting functions,
including power suppressed terms.
In the following, we will recall how to derive the collinear
splitting functions, using the algorithm of~\cite{Catani:1999ss}.
In Secs.~\ref{sec:radiation_kinematics} and~\ref{sec:splitting_kinematics}
we will then determine their correct arguments in terms of the
kinematical parameters used in the parton-shower. 

\subsection{Purely collinear splitting functions}
If two partons, $i$ and $j$, of an $n$-parton QCD amplitude become
collinear, the squared amplitude factorizes as
\begin{equation}\label{eq:collinear_factorization_fs}
    _{n}\langle1,\ldots,n|1,\ldots,n\rangle_{n}=
    \sum_{\lambda,\lambda'=\pm}
    \, _{n-1}{\Big<}1,\ldots,i\!\!\backslash(ij),\ldots,j\!\!\!\backslash,\ldots,n\Big|
    \frac{8\pi\alpha_s\,P^{\lambda\lambda'}_{(ij)i}(z)}{2p_ip_j}
    \Big|1,\ldots,i\!\!\backslash(ij),\ldots,j\!\!\!\backslash,\ldots,n\Big>_{n-1}\;,
\end{equation}
where the notation $i\!\!\backslash$ indicates that parton $i$ is
removed from the original amplitude, and where $(ij)$ is the
progenitor of partons $i$ and $j$.  The $P^{\lambda\lambda'}_{ab}(z)$
are the spin-dependent DGLAP splitting functions, which depend on the
momentum fraction $z$ of parton $i$ with respect to the mother parton,
$(ij)$, and on the helicities $\lambda$~\cite{Dokshitzer:1977sg,
  Gribov:1972ri,Lipatov:1974qm,Altarelli:1977zs,Catani:2000ef,Catani:2002hc}.

These splitting functions can be derived using the following
Sudakov parametrization of the momenta of the splitting products
\begin{equation}\label{eq:sudakov_decomposition}
  \begin{split}
    p_i^\mu=&\;%z_i\hat{p}_{ij}^\mu+\beta_i\bar{n}^\mu+k_\perp^\mu
    z_i\hat{p}_{ij}^\mu+\frac{-k_t^2}{z_i\,2p_{ij}\bar{n}}\,\bar{n}^\mu+k_t^\mu\;,\qquad&
    p_j^\mu=&\;%z_j\hat{p}_{ij}^\mu+\beta_j\bar{n}^\mu-k_\perp^\mu
    z_j\hat{p}_{ij}^\mu+\frac{-k_t^2}{z_j\,2p_{ij}\bar{n}}\,\bar{n}^\mu-k_t^\mu\;.
  \end{split}
\end{equation}
In this context, 
$\hat{p}_{ij}^\mu=p_{ij}^\mu-p_{ij}^2/(2p_{ij}\bar{n})\bar{n}^\mu$,
and $\bar{n}^\mu$ is a light-like auxiliary vector, linearly
independent of $\hat{p}_{ij}^\mu$ and $k_t^\mu$.
Equation~\eqref{eq:sudakov_decomposition} implies that we can
compute the light-cone momentum fractions, $z_i$ and $z_j$ as
\begin{equation}\label{eq:def_zi_zj}
    z_i=\frac{p_i\bar{n}}{p_{ij}\bar{n}}\,,
    \qquad\text{and}\qquad
    z_j=\frac{p_j\bar{n}}{p_{ij}\bar{n}}\;.
\end{equation}

The tree-level $g\to q\bar{q}$ and $g\to gg$ collinear splitting
functions are obtained by projecting the $\mathcal{O}(\alpha_s)$
expression for the discontinuity of the gluon propagator onto the
physical degrees of freedom of the gluon field, using the polarization
sum in a physical gauge~\cite{Catani:1999ss}.  Gauge invariance of the
underlying Born matrix element and the relation $k_t^2=-2p_ip_j\,z_iz_j$,
derived from Eq.~\eqref{eq:sudakov_decomposition}, result in the familiar
expressions
\begin{equation}\label{eq:pgq_spin}
  \begin{split}
    P^{\mu\nu}_{gq}(p_i,p_j,\bar{n})=&\;T_R\,\bigg[
    -g^{\mu\nu}+4z_iz_j\frac{k_t^\mu k_t^\nu}{k_t^2}\,\bigg]\;,\\
    P^{\mu\nu}_{gg}(p_i,p_j,\bar{n})=&\;C_A\,\bigg[
    -g^{\mu\nu}\bigg(\frac{z_i}{z_j}+\frac{z_j}{z_i}\bigg)
    -2(1-\eps)\,z_iz_j\frac{k_t^\mu k_t^\nu}{k_t^2}\,\bigg]\;.
  \end{split}
\end{equation}
The spin-averaged quark splitting function in the collinear limit can
be obtained by projecting the vertex function onto the collinear
direction~\cite{Catani:1999ss}, leading to
\begin{equation}\label{eq:coll_qqg}
  \begin{split}
    \,P_{qq}(p_i,p_j,\bar{n})
    =C_F\,\bigg[\,\frac{2z_i}{z_j}+(1-\eps)(1-z_i)\,\bigg]\;.
  \end{split}
\end{equation}
We define the difference of the full splitting functions of
Eqs.~\eqref{eq:pgq_spin} and~\eqref{eq:coll_qqg} and their
eikonal limit as the {\em purely collinear} splitting function,
$P_{\parallel}(p_i,p_j)$.  Using the known spin dependence of the
quark splitting function, we obtain the following spin-dependent and
spin-averaged expressions for final-state splittings (denoted by a
superscript ${(F)}$)
\begin{equation}\label{eq:coll_sfs_avg}
  \begin{split}
    P^{ss'\,\rm (F)}_{qq\,\parallel}(p_i,p_j,\bar{n})=&\;\delta^{ss'}
    C_F\,(1-\eps)(1-z_i)\;,&
    P^{\,\rm (F)}_{qq\,\parallel}(p_i,p_j,\bar{n})=&\;C_F\,(1-\eps)(1-z_i)\;,\\
    P^{\mu\nu\,\rm (F)}_{gg\,\parallel}(p_i,p_j,\bar{n})=&\;-2C_A\,%\bigg[
    %-g^{\mu\nu}(z_i+z_j)^2\frac{z_j}{z_i}
    (1-\eps)\,z_iz_j\frac{k_t^\mu k_t^\nu}{k_t^2}\;,\qquad &
    P^{\,\rm (F)}_{gg\,\parallel}(p_i,p_j,\bar{n})=&\;2C_A\,z_iz_j\;,\\
    P^{\mu\nu\,\rm (F)}_{gq\,\parallel}(p_i,p_j,\bar{n})=&\;P^{\mu\nu}_{gq}(p_i,p_j)\;,&
    P^{\,\rm (F)}_{gq\,\parallel}(p_i,p_j,\bar{n})=&\;T_R\,\bigg[\;1-\frac{2\,z_iz_j}{1-\eps}\;\bigg]\;.
  \end{split}
\end{equation}
At this point we would like to stress that $z_i$ and $z_j$ depend on
the precise form of the momentum mapping, and that they are not
necessarily identical to the parton-shower splitting variables $z$ and
$1-z$.  This has implications in particular for the splitting
functions in initial-state evolution and will be discussed in
Secs.~\ref{sec:radiation_kinematics}
and~\ref{sec:splitting_kinematics}.

Crossing parton $i$ into the initial state, we obtain the following
collinear factorization formula
\begin{equation}\label{eq:collinear_factorization_is}
    _{n}\langle1,\ldots,n|1,\ldots,n\rangle_{n}=
    \sum_{\lambda,\lambda'=\pm}
    \, _{n-1}{\Big<}1,\ldots,i\!\!\backslash(ij),\ldots,j\!\!\!\backslash,\ldots,n\Big|
    \frac{8\pi\alpha_s\,P^{\lambda\lambda'}_{(ij)i}(x)}{2p_ip_j\,x}
    \Big|1,\ldots,i\!\!\backslash(ij),\ldots,j\!\!\!\backslash,\ldots,n\Big>_{n-1}\;,
\end{equation}
where $x=1/z$ is the momentum fraction of parton $(ij)$ with
respect to the initial-state parton $i$. 
Equation~\eqref{eq:collinear_factorization_fs} is obtained from
Eq.~\eqref{eq:collinear_factorization_fs} via the crossing relation
$P_{ab}(1/x)=-P_{ab}(x)/x$~\cite{Gribov:1972ri,Gribov:1972rt}.
The splitting functions $P^{\lambda\lambda'}_{ab}(x)$ are therefore
determined by Eqs.~\eqref{eq:pgq_spin} and~\eqref{eq:coll_qqg}.
However, the matching to the soft radiation pattern differs for
initial-state splittings, because an initial-state particle of
vanishing energy will lead to a vanishing cross section
(see for example Sec.5.4 in~\cite{Catani:1996vz}). This leads
to the following expressions for the flavor-diagonal splitting functions
in the initial state (denoted by a superscript ${(I)}$)
\begin{equation}\label{eq:pgg_spin}
    P^{\rm (I)}_{qq\,\parallel}(p_i,p_j,\bar{n})=P_{qq}(p_i,p_j,\bar{n})\;,\qquad
    P^{\rm (I)}_{gg\,\parallel}(p_i,p_j,\bar{n})=C_A\,\bigg[
    -g^{\mu\nu}x_ix_j
    +2(1-\eps)\,\frac{x_j}{x_i}\frac{k_t^\mu k_t^\nu}{k_t^2}\,\bigg]\;.
\end{equation}
All other purely collinear splitting functions remain the same.
We have simplified the notation by defining $x_i=1/z_i$ and $x_j=-x_i z_j$.
We stress again that differences in the purely collinear components of
the spin-averaged DGLAP splitting functions can arise from the fact
that $x_i$ may not be equal to $x$, where $x$ is the initial-state
parton shower splitting variable.  In practical applications, this
typically leads to a suppression of $1/x$ enhanced parton splittings
at large transverse momenta.  We will return to this question in
Secs.~\ref{sec:radiation_kinematics} and~\ref{sec:splitting_kinematics},
see in particular the discussion following Eq.~\eqref{eq:def_zeta_xi_is}.

\subsection{Soft limit and soft-collinear matching}
In the limit that gluon $j$ becomes soft, the squared amplitude
factorizes as~\cite{Bassetto:1984ik}
\begin{equation}\label{eq:soft_factorization}
    _{n}\langle1,\ldots,n|1,\ldots,n\rangle_{n}=-8\pi\alpha_s\sum_{i,k\neq i,j}
    \,_{n-1}\big<1,\ldots,j\!\!\!\backslash,\ldots,n\big|{\bf T}_i{\bf T}_k\,w_{ik,j}
    \big|1,\ldots,j\!\!\!\backslash,\ldots,n\big>_{n-1}\;,
\end{equation}
where ${\bf T}_i$ and ${\bf T}_k$ are the color insertion operators
defined in~\cite{Bassetto:1984ik,Catani:1996vz}.  In the \Alaric
parton-shower algorithm~\cite{Herren:2022jej}, the eikonal factor
$w_{ik,j}$ is split into an angular radiator $W_{ik,j}$ and the gluon
energy according to $w_{ik,j}=W_{ik,j}/E_j^2$. The angular radiator
function
\begin{equation}\label{eq:eikonal_split}
  W_{ik,j}=\frac{1-\cos{\theta_{ik}}}{(1-\cos{\theta_{ij}})(1-\cos{\theta_{jk}})}
\end{equation}
is matched to the collinear splitting functions by partial fractioning:
\begin{equation}\label{eq:partfrac_soft_matching}
  W_{ik,j}=\bar{W}_{ik,j}^i+\bar{W}_{ki,j}^k\;,
  \qquad\text{where}\qquad
  \bar{W}_{ik,j}^i=\frac{1-\cos\theta_{jk}}{
      2-\cos\theta_{ij}-\cos\theta_{jk}}\,
    W_{ik,j}\;.
\end{equation}
In the collinear limit for partons $i$ and $j$, the eikonal factor
$w_{ik,j}$ can be identified with the eikonal term of the DGLAP
splitting functions $P_{aa}(z)$.  Matching the soft to the collinear
splitting functions in the improved large-$N_c$ limit is achieved
by replacing
\begin{equation}\label{eq:matching_soft_coll_fs}
  \frac{P^{\,\rm (F)}_{(ij)i}(p_i,p_j,\bar{n})}{2p_ip_j}\to
  \frac{{\bf T}_{ij}^2}{N_{{\rm spec}}}\,
  \sum_{k\in {\rm specs}}^{N_{{\rm spec}}}
  \bigg[\,\delta_{(ij)i}\frac{\bar{W}_{ik,j}^i}{E_j^2}
    +\delta_{(ij)j}\frac{\bar{W}_{jk,i}^j}{E_i^2}\,\bigg]+
  \frac{P^{\,\rm (F)}_{(ij)i\,\parallel}(p_i,p_j,\bar{n})}{2p_ip_j}\;,
\end{equation}
where the sum runs over all color-connected partons, and $N_{\rm
  spec}$ stands for the number of color spectators.  While
initial-state parton evolution must respect Gribov-Lipatov
reciprocity~\cite{Gribov:1972ri,Gribov:1972rt}, we need to take into
account that the amplitude cannot develop a soft singularity in the
initial-state momentum.  Therefore,
\begin{equation}\label{eq:matching_soft_coll_is}
  \frac{P^{\,\rm (I)}_{i(ij)}(p_i,p_j,\bar{n})}{2p_ip_j\,x}\to
  \delta_{i(ij)}\,\frac{{\bf T}_{ij}^2}{N_{\rm spec}}\,
  \sum_{k\in {\rm specs}}^{N_{\rm spec}}\frac{\bar{W}_{ik,j}^i}{E_j^2}+
  \frac{P^{\,\rm (I)}_{i(ij)\,\parallel}(p_i,p_j,\bar{n})}{2p_ip_j\,x}\;.
\end{equation}
The two soft contributions to the gluon splitting function are treated
as two different radiators~\cite{Hoche:2015sya}.  The soft matching
introduces a dependence of the splitting functions on the color
spectators, $k$, and their momenta define directions independent of
$\hat{p}_{ij}$~\cite{Herren:2022jej}.

\section{Momentum mapping}
\label{sec:kinematics}
Parton shower algorithms are based on the notion of adding additional
partons to an already existing ensemble of particles, while
maintaining four-momentum conservation and on-shell conditions.  This
procedure requires a method to map the momenta of the Born process to
a kinematical configuration after emission.  The mappings are linked
to the factorization of the differential phase-space element for a
multi-parton configuration. Collinear safety a basic requirement
for their construction.  In addition, a mapping is NLL-safe if it
preserves the topological features of previous
radiation~\cite{Dasgupta:2018nvj,Dasgupta:2020fwr}.
Since the momentum mapping in most modern parton showers has been
identified as the main stumbling block to achieving next-to-leading
logarithmic precision, we will begin the description of \Alaric's
initial-state evolution algorithm by discussing the kinematics.

\subsection{Soft radiation kinematics}
\label{sec:radiation_kinematics}
\begin{figure}[t]
\includegraphics[width=\textwidth]{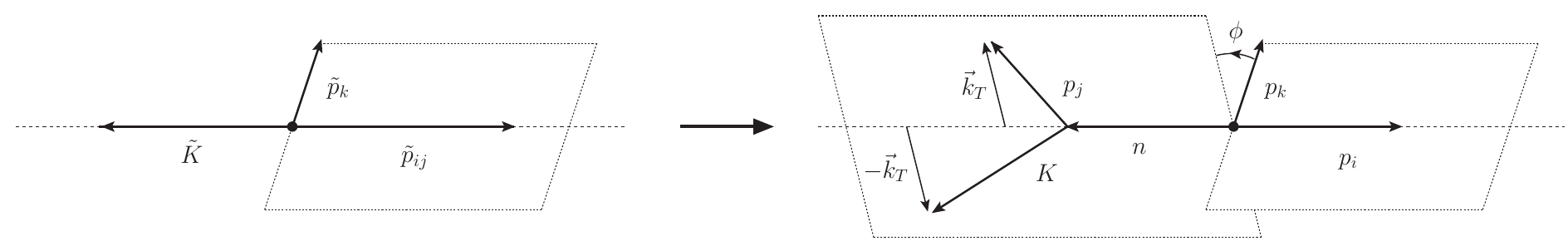}
\caption{
Sketch of the momentum mapping for soft radiation and initial-state
splittings.  All momenta are considered outgoing.  Note that $p_k$
only acts as a reference for the definition of the azimuthal angle
$\phi$.  See the main text for details.
\label{fig:kinematics_ff_rad}}
\end{figure}
This section details the algorithm for the construction of momenta in
soft emissions. The momentum mapping is sketched in
Fig.~\ref{fig:kinematics_ff_rad}.  We identify the splitter momentum,
$\tilde{p}_i$, and define a recoil momentum, $\tilde{K}$.  In contrast
to conventional dipole-like parton showers where the recoil momentum
is usually given by the color spectator, in \Alaric this momentum can
be chosen freely, with the condition that it must provide a hard
scale.  In most practical applications we will define $\tilde{K}$ as
the sum of all final-state momenta (in the case of final-state
branchings also including the momentum of the splitting particle).
Together, the momenta $\tilde{K}$ and $\tilde{p}_i$ define the
reference frame of the splitting.  The momentum of the color
spectator, $\tilde{p}_k$, defines an additional direction, and
provides the reference for the azimuthal angle, $\phi$.  To obtain the
momenta after emission, the emitter is scaled by a factor $z$, and the
emitted momentum, $p_j$, is constructed with transverse momentum
component $\vec{k}_T$ and suitable light-cone momenta.  The recoil is
absorbed by all particles that constitute the recoil momentum
$\tilde{K}$.  To parametrize the splitting kinematics, we make use of
some of the notation in~\cite{Catani:1996vz,Herren:2022jej}, in
particular
\begin{equation}\label{eq:def_v_z_cs}
  v=\frac{p_ip_j}{\tilde{p}_i\tilde{K}}
  \qquad\mathrm{and}\qquad
  z=\frac{p_i\tilde{K}}{\tilde{p}_i\tilde{K}}\;.
\end{equation}
The momentum mapping for emitter $\tilde{p}_i$ and recoil momentum
$\tilde{K}$ is fixed by
\begin{equation}\label{eq:fi_emit_spec}
  \begin{split}
    p_i=&\;\,z\,\tilde{p}_i\;,\\
    p_j=&\;\,(1-z)\,\tilde{p}_i+v\big(\tilde{K}-(1-z+2\kappa)\,\tilde{p}_i\big)-k_\perp\;,\\
    K=&\;\tilde{K}-v\big(\tilde{K}-(1-z+2\kappa)\,\tilde{p}_i\big)+k_\perp\;,
  \end{split}
\end{equation}
with the absolute value of the transverse momentum given by
\begin{equation}
  {\rm k}_\perp^2=v(1-v)(1-z)\,2\tilde{p}_i\tilde{K}-v^2\tilde{K}^2\;.
\end{equation}
For initial-state splitters, the energy fraction $z$ is replaced by
$1/x$.  If the momentum $\tilde{K}$ is composed of the two
initial-state momenta, all final-state momenta are subjected to a
Lorentz transformation
\begin{equation}\label{eq:lorentz_trafo_fs}
  p_l^\mu\to\Lambda^\mu_{\;\nu}(K,\tilde{K}) \,p_l^\nu\;,  
  \qquad\text{where}\qquad
  \Lambda^\mu_{\;\nu}(\tilde{K},K)=g^\mu_{\;\nu}
  -\frac{2(K+\tilde{K})^\mu(K+\tilde{K})_\nu}{(K+\tilde{K})^2}
  +\frac{2K^\mu \tilde{K}_\nu}{\tilde{K}^2}\;.
\end{equation}
If the momentum $\tilde{K}$ is composed of final-state momenta, those
momenta are subjected to a Lorentz transformation
$p_l^\mu\to\Lambda^\mu_{\;\nu}(\tilde{K},K)\,p_l^\nu$, with
$\Lambda^\mu_{\;\nu}(\tilde{K},K)$ given by
Eq.~\eqref{eq:lorentz_trafo_fs}.

It remains to determine the variables $z_i$ and $z_j$ in
Sec.~\ref{sec:splitting_functions}, which are needed to evaluate the
purely collinear splitting functions.  Expanding Eq.~\eqref{eq:fi_emit_spec}
in terms of the large forward light-cone momentum,
$\hat{p}_{ij}^\mu=p_{ij}^\mu-p_{ij}^2/(2p_{ij}\bar{n})\bar{n}^\mu$,
the small transverse components, and the very small anti-collinear
components, we obtain
\begin{equation}\label{eq:boosted_momenta_radkin_coll}
  \begin{split}
    p_{i}=&\;\frac{z}{1-v(1-z+\kappa)}\,\hat{p}_{ij}
    +\frac{z}{1-v(1-z+\kappa)}\,k_\perp
    +\mathcal{O}\bigg(\frac{k_\perp^2}{2\tilde{p}_i\tilde{K}}\bigg)\;,\\
    p_{j}=&\;\frac{(1-z)(1-v)-v\kappa}{1-v(1-z+\kappa)}\,\hat{p}_{ij}
    -\frac{z}{1-v(1-z+\kappa)}\,k_\perp
    +\mathcal{O}\bigg(\frac{k_\perp^2}{2\tilde{p}_i\tilde{K}}\bigg)\;.\\
  \end{split}
\end{equation}
Having obtained an expression equivalent to
Eq.~\eqref{eq:sudakov_decomposition}, it is apparent that the
momentum fractions that appear in the purely collinear splitting
functions, Eqs.~\eqref{eq:coll_sfs_avg} and~\eqref{eq:pgg_spin} are
given by
\begin{equation}\label{eq:def_zeta_xi_fs}
  \begin{split}
    z_i=&\;\frac{z}{1-v(1-z+\kappa)}\;,\qquad
    &z_j=&\;1-\frac{z}{1-v(1-z+\kappa)}\;.
  \end{split}
\end{equation}
In initial-state evolution, the replacements $z\to 1/x$, $z_i\to
1/x_i$ and $z_j\to -x_j/x_i$ change Eq.~\eqref{eq:def_zeta_xi_fs} to
\begin{equation}\label{eq:def_zeta_xi_is}
  \begin{split}
    x_i=&\;x+v-v\,x\,(1+\kappa)\;,\qquad
    &x_j=&\;1-x-v+v\,x\,(1+\kappa)\;.
  \end{split}
\end{equation}
In addition, the transverse momentum $k_t^\mu$ appearing in the Sudakov
decomposition Eq.~\eqref{eq:sudakov_decomposition}, and hence in the
spin-dependent splitting functions in Sec.~\ref{sec:splitting_functions},
is expressed in terms of the radiation kinematics variables appearing
in Eq.~\eqref{eq:fi_emit_spec} as $z_i\,k_\perp^\mu$~\cite{Catani:1996vz}.
Note that for initial-state emissions with the spectator being the
complete final state, Eq.~\eqref{eq:def_zeta_xi_is} simplifies to
$x_i=x+v$ and $x_j=1-x-v$. This relation has been used in the context
of a Catani-Seymour dipole shower in Refs.~\cite{Jones:2017giv,ATLAS:2021yza}
to obtain an improved approximation of the splitting functions and generally
leads to a reduction of emission probabilities from terms in the splitting
functions that are proportional to $1/x_i$.

\subsection{Collinear splitting kinematics}
\label{sec:splitting_kinematics}
\begin{figure}[t]
\includegraphics[width=\textwidth]{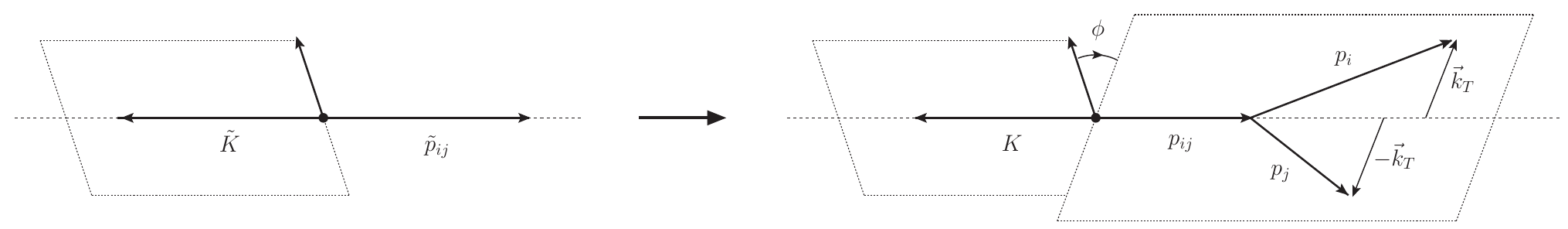}
\caption{
Sketch of the momentum mapping for collinear radiation.  All momenta
are considered outgoing.  Note that, again, $p_k$ only acts as a
reference for the definition of the azimuthal angle $\phi$.  See the
main text for details.
\label{fig:kinematics_ff_split}}
\end{figure}
As discussed in Sec.~\eqref{sec:splitting_functions}, collinear parton
evolution is easier understood in a phase-space parametrization where
the splitting products compensate each others transverse recoil with
respect to the direction of the progenitor.  For the implementation of
the purely collinear components of final-state splitting functions we
therefore choose a kinematics mapping that is closely related
to~\cite{Hoche:2020pxj,Nagy:2020rmk}.  It has been
shown~\cite{Nagy:2020rmk} that this type of mapping satisfies the
criteria for NLL precision if it is applied to the purely collinear
splitting functions only.  The proof rests on similar arguments as the
proof of accuracy for the radiation kinematics of
Sec.~\ref{sec:radiation_kinematics}~\cite{Herren:2022jej}.

The momentum mapping is sketched in
Fig.~\ref{fig:kinematics_ff_split}.  We identify the splitter
momentum, $\tilde{p}_i$, and define a longitudinal recoil momentum,
$\tilde{K}$.  Again, this recoil momentum can be freely defined.  In
most practical applications we use the sum of final-state momenta,
excluding the momentum of the splitter.  Together, the momenta
$\tilde{K}$ and $\tilde{p}_i$ define the reference frame of the
splitting.  And, as before, the momentum of the color spectator,
$\tilde{p}_k$, defines an additional direction, which provides the
reference for the azimuthal angle, $\phi$.  To obtain the momenta
after emission, we invoke the massive splitting kinematics
of~\cite{Catani:2002hc}, in which the emitter is scaled and the
momentum $K$ absorbs the longitudinal recoil, while the transverse
recoil is compensated locally between the splitting products.  We make
use of some of the notation in~\cite{Catani:1996vz}, in particular
\begin{equation}\label{eq:def_y_z_cdst}
  y=\frac{p_ip_j}{p_ip_j+(p_i+p_j)K}
  \qquad\mathrm{and}\qquad
  z=\frac{p_iK}{(p_i+p_j)K}\;,
\end{equation}
and we define $\kappa=K^2/(2\tilde{p}_{ij}\tilde{K})$.
In terms of the additional variables
\begin{equation}
  \begin{split}
    \zeta=&\;\frac{1+y-\sqrt{(1-y)^2-4y\kappa}}{2y(1+\kappa)}\;,\qquad
    &\bar{z}=&\;\frac{\displaystyle z\big(1-y\big)\zeta(1-\zeta y)
    -\zeta^2y\kappa}{\displaystyle
    (1-\zeta y)^2-\zeta^2y\kappa}\;,
  \end{split}
\end{equation}
the momenta after the splitting are given by
\begin{equation}\label{eq:coll_split_ff}
  \begin{split}
    p_i^\mu=&\;\bar{z}\;
    \frac{\tilde{p}_{ij}^\mu}{\zeta}\,
    +(1-\bar{z})\,y
    \zeta\big(\tilde{K}^\mu-\kappa\,\tilde{p}_{ij}^\mu\big)
    +k_\perp^\mu\;,\\
    p_j^\mu=&\;(1-\bar{z})\,
    \frac{\tilde{p}_{ij}^\mu}{\zeta}\,
    +\bar{z}\,y\zeta\big(\tilde{K}^\mu-\kappa\,\tilde{p}_{ij}^\mu\big)
    -k_\perp^\mu\;,\\
    K^\mu=&\;\bigg(1-\frac{1-y\kappa\zeta^2}{\zeta}\bigg)\,\tilde{p}_{ij}^\mu
    +(1-y\zeta)\tilde{K}^\mu\;.
  \end{split}
\end{equation}
The transverse momentum squared is given by
\begin{equation}\label{eq:kt2_split}
  {\rm k}_\perp^2=y\bar{z}(1-\bar{z})\,2\tilde{p}_{ij}\tilde{K}\;.
\end{equation}
This particular scheme cannot be used in initial-state splittings,
because the momentum of the splitter and at least one of the splitting
products must be aligned. We therefore use the soft radiation kinematics
also for the purely collinear initial-state splittings. If the momentum $\tilde{K}$
was constructed from multiple final-state momenta, those momenta are subjected to a
Lorentz transformation $p_l^\mu\to\Lambda^\mu_{\;\nu}(\tilde{K},K)\,p_l^\nu$,
with $\Lambda^\mu_{\;\nu}(\tilde{K},K)$ given by Eq.~\eqref{eq:lorentz_trafo_fs}.

Equation~\eqref{eq:coll_split_ff} has the form of
Eq.~\eqref{eq:sudakov_decomposition}, and we can read off the
momentum fractions that appear in the purely collinear splitting
functions, Eqs.~\eqref{eq:coll_sfs_avg} and~\eqref{eq:pgg_spin}:
\begin{equation}\label{eq:def_zeta_xi_split}
  \begin{split}
    z_i=&\;\bar{z}
    \qquad\mathrm{and}\qquad
    &z_j=&\;1-\bar{z}
  \end{split}
\end{equation}
with the transverse momentum squared, $k_t^2$, in collinear splitting
kinematics given by Eq.~\eqref{eq:kt2_split}. Note that this relation
differs from the definition $z_i=z$ in a Catani-Seymour dipole like
final state parton shower only due to the mass of the recoil momentum.
In particular, it is identical if $\kappa=0$, as for example in the first
emission in $e^+e^-\to$hadrons, or in pure jet production at hadron
colliders.

\section{The evolution algorithm}
\label{sec:evolution}
For initial-state evolution, we use the original definition of the
evolution variable in~\cite{Herren:2022jej}.  It is related to the
energies and polar angle in the rest frame of $n$, where we have the
simple relations
\begin{equation}\label{eq:energies_n_frame_fs}
    E_i=z\,\frac{\tilde{p}_i\tilde{K}}{\sqrt{n^2}}\;,
    \qquad
    E_j=E_i\,\frac{1-z}{z}\;,
    \qquad\text{and}\qquad
    n^2=2\tilde{p}_i\tilde{K}\,(1-z+\kappa)\;.
\end{equation}
The polar angle, $\theta_j$, of the emission is given by
\begin{equation}
\label{eq:ct_nframe_fs}
    1-\cos\theta_j^{\,i}=2v\,\frac{1-z+\kappa}{1-z}\;.
\end{equation}
In terms of these quantities, the initial-state evolution
variable is defined as (see Eq.~(41) of~\cite{Herren:2022jej})
\begin{equation}\label{eq:def_t_n}
    t^{(n)}=2E_j^2\,(1-\cos\theta_j^{\;i})
    =v\,(1-z)\,2\tilde{p}_i\tilde{K}\;.
\end{equation}
The same definition can also be used for final-state evolution.
However, we also introduce a variant of the original proposal, which
will become our default choice: We determine the evolution variable using 
energies and angles in the rest frame of the recoil momentum, $K$, after
the emission.  In the soft limit, $p_j\to 0$, this frame coincides
with the frame defined by $n$.  The energies of particles $i$ and $j$
in the $K$-frame are given by
\begin{equation}
    E_i=(1-v)\frac{z\,\tilde{p}_i\tilde{K}}{\sqrt{\tilde{K}^2}}\;,\qquad
    E_j=\frac{E_i}{1-v}\,\frac{1-z}{z}\;.
\end{equation}
The polar angle $\theta_j^{\,i}$ of the emission is determined by
\begin{equation}
    1-\cos\theta_j^{\,i}=\frac{2\kappa}{1-z}\frac{v}{1-v}\;.
\end{equation}
The final-state evolution variable of the parton shower
in this scheme is defined as
\begin{equation}\label{eq:def_t_K}
    t^{(K)}=2E_j^2\,(1-\cos\theta_j^{\;i})
    =\frac{v}{1-v}\,(1-z)\,2\tilde{p}_i\tilde{K}\;.
\end{equation}
The advantage of the redefinition is a simplified determination
of the upper bound on the evolution variable.  For example, in color
singlet decays one has $t^{(K)}\le\sqrt{\tilde{K}^2}$.  The Jacobian
factor for the transformation $\ln v\to\ln t$ is given by
$(1-z)/(1-z+\tau)$.  We will investigate the numerical effect of
different choices for the evolution scheme in Sec.~\ref{sec:results}.

\section{Multi-jet Merging}
\label{sec:merging}
The physics modeling of parton-shower simulations can be improved
systematically with the help of multi-jet merging~\cite{Catani:2001cc,
  Lonnblad:2001iq,Krauss:2002up,Alwall:2007fs,Hoeche:2009rj,
  Lonnblad:2011xx,Gehrmann:2012yg,Hoeche:2012yf,Frederix:2012ps,
  Lonnblad:2012ix,Platzer:2012bs}.
This is achieved by including higher multiplicity tree-level fixed
order calculations with well separated parton-level jets, while
maintaining both the logarithmic accuracy of the parton shower and the
fixed order accuracy.  Here, we implement the leading-order merging
method described in~\cite{Hoeche:2009rj}, which can be described as
follows:
\begin{enumerate}
\item The phase space of parton-shower emissions is restricted to the
  complement of the phase space of the fixed-order calculations.  For
  example, in the combination of $pp\to Z$ and $pp\to Zj$, with
  $p_{T,j}\ge p_{T,\rm cut}$, the phase space of the first
  parton-shower emission would be restricted to $p_\perp<p_{\perp,\rm
    cut}$.  This is called the jet veto, the variable used to separate
  the phase space is called the jet criterion, and the separation
  scale is called the merging scale, $Q_{\rm cut}$.
\item The fixed-order result is modified to include higher-order
  corrections as resummed in the parton-shower approach.  This
  procedure consists of multiple steps:
  \begin{enumerate}
  \item Re-interpreting the final-state configuration of the
    fixed-order calculation as having originated from a parton
    cascade~\cite{Andre:1997vh}.  This is called {\it clustering}, and
    the representations of the final-state configuration in terms of
    parton branchings are called parton-shower histories.
  \item Choosing appropriate scales for evaluating the strong coupling
    at each branch point in the cascade, thereby resumming
    higher-order corrections to soft-gluon
    radiation~\cite{Amati:1980ch,Catani:1990rr}.  This procedure is
    called $\alpha_s$-reweighting.
  \item Weighting by appropriate no-emission probabilities,
    representing the resummed unresolved real and virtual
    corrections~\cite{Catani:2001cc}.  This procedure is called
    Sudakov reweighting. It is implemented using pseudo
    showers~\cite{Lonnblad:2001iq}.
  \end{enumerate}
\end{enumerate}
The jet clustering procedure for the \Alaric parton shower requires
extra care, because multiple histories (soft and collinear) may exist
for each combination of external partons.  Due to the difference
between radiation and splitting kinematics, they differ not only in
their associated weight, but also in the kinematics of the underlying
Born state.

\section{Numerical Results}
\label{sec:results}
In this section we present first numerical results obtained with the
\Alaric parton shower for hadron colliders, as implemented in the event
generation framework \Sherpa~\cite{Gleisberg:2003xi,Gleisberg:2008ta,
  Sherpa:2019gpd}.  We set $C_F=(N_c^2-1)/(2N_c)=4/3$ and $C_A=3$,
all quarks are considered as massless, but we implement flavor thresholds
at $m_c=1.42$~GeV and $m_b=4.92$~GeV.  The running coupling is evaluated at
two loop accuracy with $\alpha_s(m_z)=0.118$.  Following standard practice
to improve the logarithmic accuracy of the parton shower, we employ the
CMW scheme~\cite{Catani:1990rr}, {\it i.e.}\ the soft eikonal
contribution to the flavor conserving splitting functions is rescaled
by $1+\alpha_s(t)/(2\pi) K$, with
$K=(67/18-\pi^2/6)\,C_A-10/9\,T_R\,n_f$.  Where appropriate, our
results include multi-jet merging at leading-order accuracy.  All
analyses are performed with Rivet~\cite{Buckley:2010ar}.

\subsection{Drell-Yan lepton pair production}
\begin{figure}[t]
  \centering
  \includegraphics[width=7.5cm]{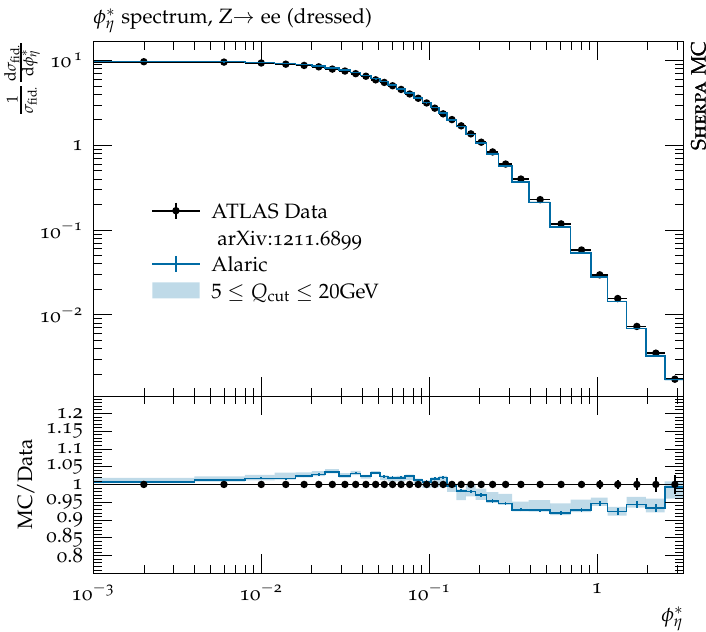}
  \includegraphics[width=7.5cm]{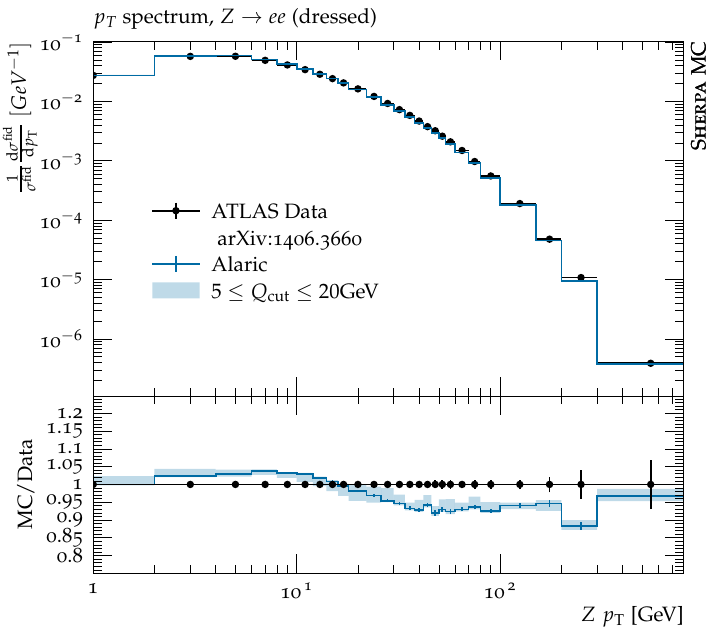}\\
  \begin{minipage}{7.5cm}
    \includegraphics[width=\textwidth]{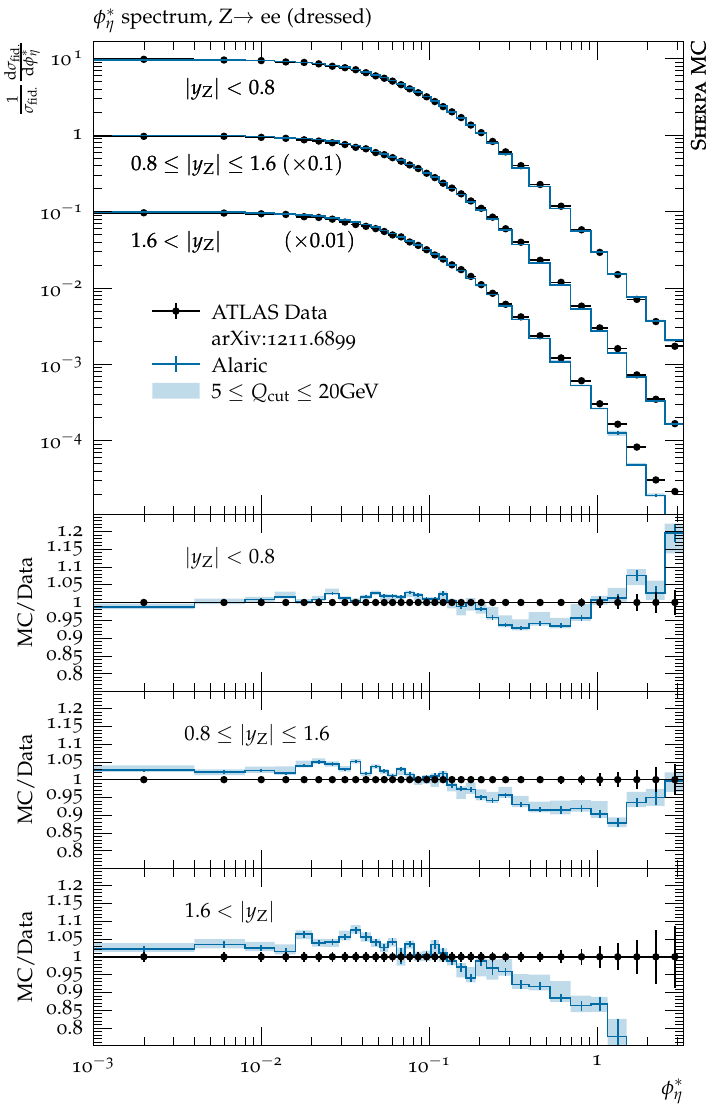}
  \end{minipage}
  \begin{minipage}{7.5cm}
    \includegraphics[width=\textwidth]{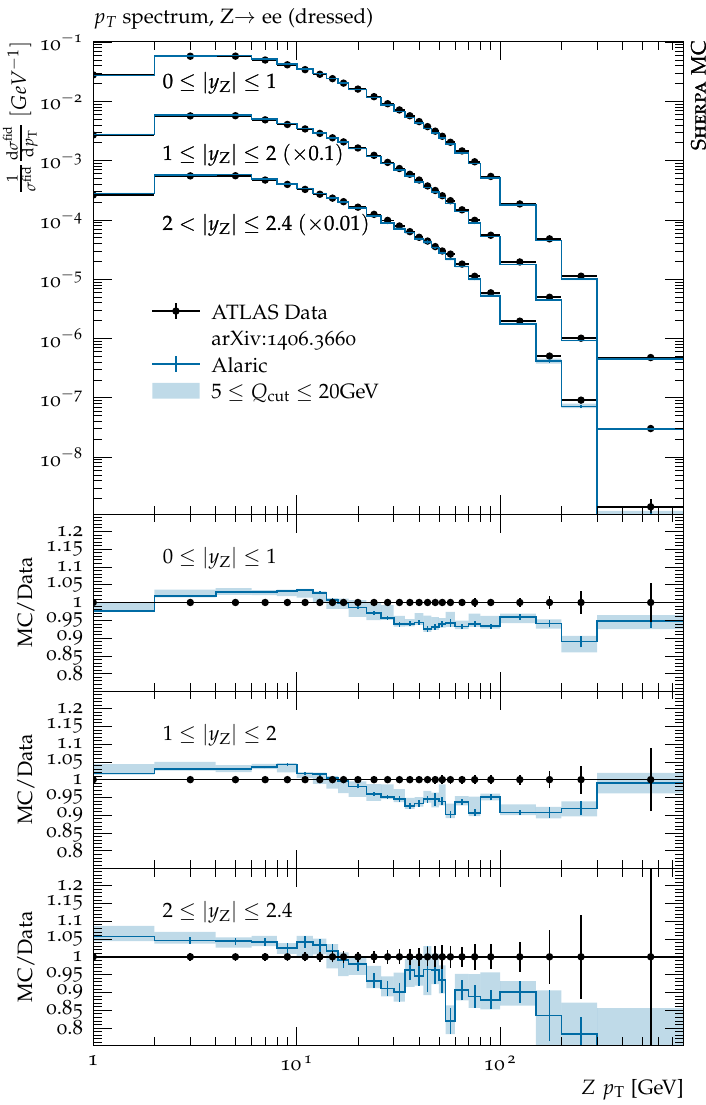}
  \end{minipage}
  \caption{
  \Alaric ME+PS merged predictions with up to three jets for
  $\phi^*_\eta$ (left) and the $Z$-boson $p_T$ (right), inclusive over
  all accessible $Z$-boson rapidities (upper panel), or in different
  bins of $|y_Z|$.  Uncertainties related to the choice of merging cut
  $Q_{\rm cut}$ are indicated with the light-blue band around the
  central value; the main plots show the overall prediction in
  comparison to the ATLAS data at 7 TeV c.m.-energy
  from~\cite{Aad:2012wfa} and~\cite{Aad:2014xaa}, while respective
  lower panels show deviations.
    \label{fig:lhc_ptspectra}}
\end{figure}
Figure \ref{fig:lhc_ptspectra} shows the transverse momentum spectrum
of the Drell-Yan lepton pair, and the angular variable
$\phi^*$~\cite{D0:2010qhp}, as predicted by a multi-jet merged
calculation with \Alaric, in comparison to experimental data from the
ATLAS collaboration~\cite{Aad:2012wfa,Aad:2014xaa}.  In this analysis,
leptons are required to have $|\eta|< 2.4$ and $p_T>20$~GeV, in
addition to the invariant mass constraint 66~GeV$\le
m_{ll}\le$116~GeV.  The leptons are dressed, {\it i.e.}, they are
combined with photons within a cone of radius $R=0.1$ .  The
uncertainty band in Fig.~\ref{fig:lhc_ptspectra} corresponds to the
variation of the merging cut between 5~GeV and 20~GeV.  In general, we
find agreement with experimental data to the level that it can be
expected from a parton-shower simulation without NLO multi-jet
merging, see for example~\cite{Hoeche:2012yf}.  Apart from very
forward regions in $Z$-boson rapidity, the deviations from data reach
at most five to ten percent.

\begin{figure}[t]
  \centering
    \includegraphics[width=7.5cm]{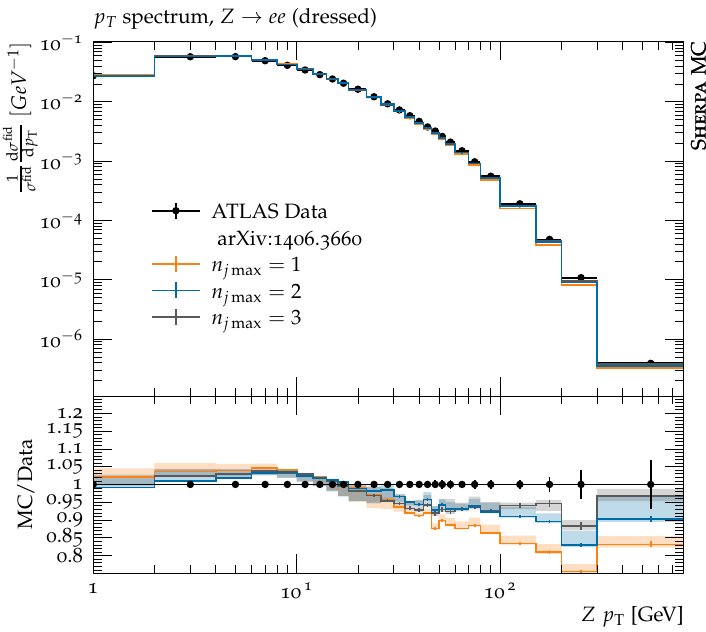}
    \includegraphics[width=7.5cm]{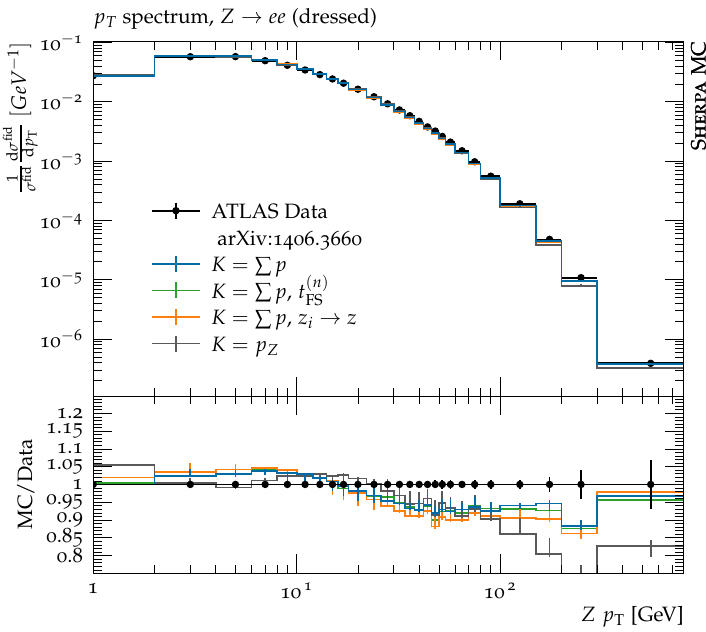}
  \caption{
  Systematic uncertainties of parton-shower predictions from \Alaric
  due to different maximal number, $n_{j,\rm max}$ of jet from
  leading-order matrix elements (left) and due to different choices
  of the recoil momentum $\tilde{K}$ (right).  
  See Fig.~\ref{fig:lhc_ptspectra} and the main text for details.
    \label{fig:dy_uncertainties}}
\end{figure}
\begin{figure}[t]
  \centering
  \includegraphics[width=7.5cm]{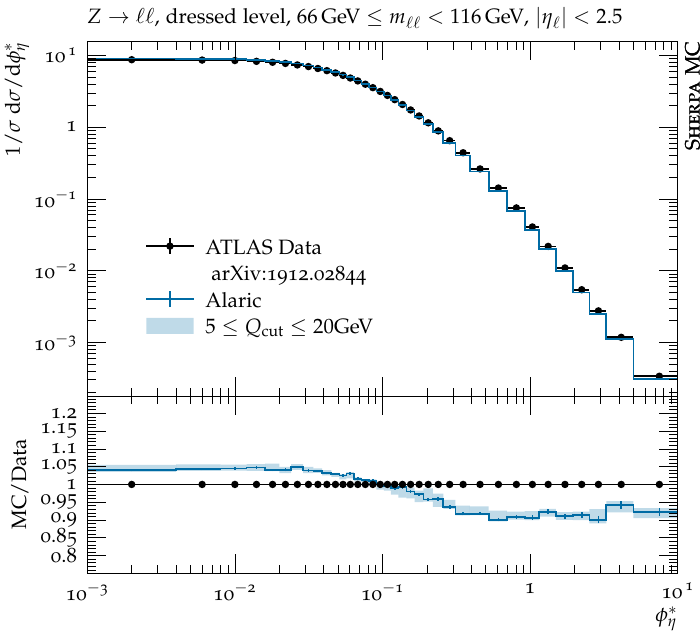}
  \includegraphics[width=7.5cm]{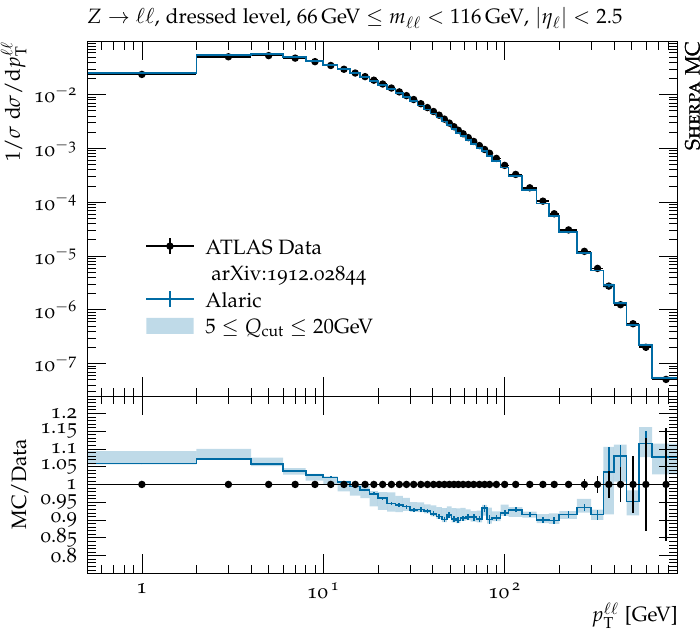}
  \caption{\Alaric ME+PS merged predictions in comparison to ATLAS data from~\cite{ATLAS:2019zci}. 
    \label{fig:dy_pt_13tev}}
\end{figure}
We include up to three jets in this simulation, but we note that the
prediction stabilizes upon including the second jet, {\it cf.} the
left panel of Fig.~\ref{fig:dy_uncertainties}.  There, we display a
variation of results with the highest jet multiplicity, $n_{j\,\rm
  max}$, the maximal number of jets described by fixed-order
calculations in the multi-jet merging.  We find that with increasing
$n_{j\,\rm max}$ the high transverse momentum region is better
described by the simulation.  This effect has been discussed in great
detail in the original literature on multi-jet
merging~\cite{Catani:2001cc,Lonnblad:2001iq,Krauss:2002up,Alwall:2007fs}.
The saturation of this effect at $n_{j\,\rm max}=2$ can be understood
by noticing that the addition of a first and second jet adds new
partonic initial state channels.

The right panel of Fig.~\ref{fig:dy_uncertainties} shows some of the
systematic uncertainties associated with the parton-shower prediction
itself.  We compare two different definitions of $K$, one where the
recoil is absorbed by the Drell-Yan lepton pair (labeled $K=p_z$), and
one where the recoil is absorbed by the complete final state (our
default choice, labeled $K=\sum p$).  While the first definition leads
to a somewhat better description of the transverse momentum spectrum
in the bulk of the distribution, it fails in the high-$p_T$ tails.
This is expected, because in the high transverse momentum region, the
invariant mass of the Drell-Yan lepton pair no longer provides the
highest scale in the process.  We also compare to a simulation where
the momentum fractions $z_i$ defined in
Eq.~\eqref{eq:sudakov_decomposition} are replaced by the splitting
variable $z$ (labeled $K=\sum p$, $z_i\to z$).  The differences
compared to the default simulation are small, because the observable
is largely insensitive to $1/x$ enhanced terms in the parton-shower
splitting functions.  Finally, we compare to a simulation where the
new evolution scheme, described in Sec.~\ref{sec:evolution}, is
replaced by the original proposal in~\cite{Herren:2022jej} (labeled
$K=\Sigma p$, $t_{\rm FS}^{(n)}$.  Again, the differences between 
the two options are small.

Figure~\ref{fig:dy_pt_13tev} displays predictions from \Alaric in
comparison to experimental data at 13 TeV c.m.-energy from the ATLAS
collaboration reported in~\cite{ATLAS:2019zci}.  In this analysis,
leptons are required to satisfy slightly different cuts, {\it i.e.}
$|\eta|< 2.5$ and $p_T>27$~GeV, in addition to the invariant mass
constraint 66~GeV$\le m_{ll}\le$116~GeV.  As before, the leptons have
been dressed with a cone of radius $R=0.1$.  By far and large we
observe similar features as in the comparison to 7~TeV data in the
upper panel of Fig.~\ref{fig:dy_uncertainties}.

\begin{figure}[t]
  \centering
  \includegraphics[width=7.5cm]{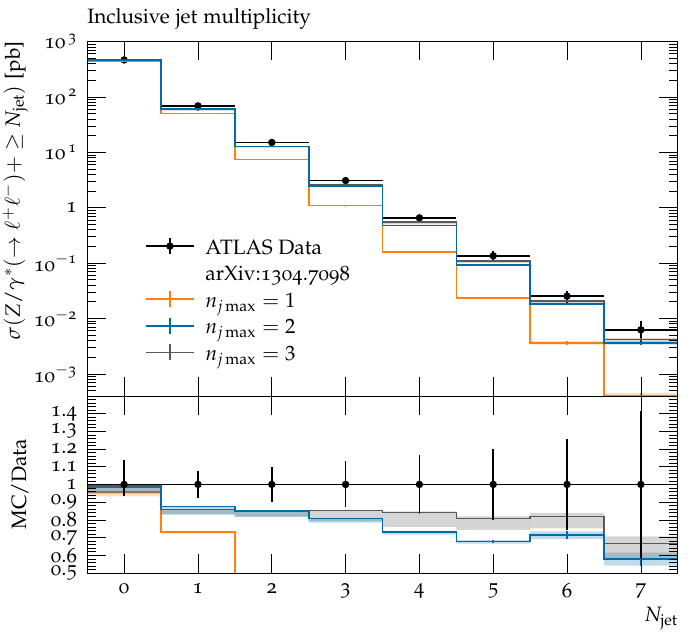}
  \includegraphics[width=7.5cm]{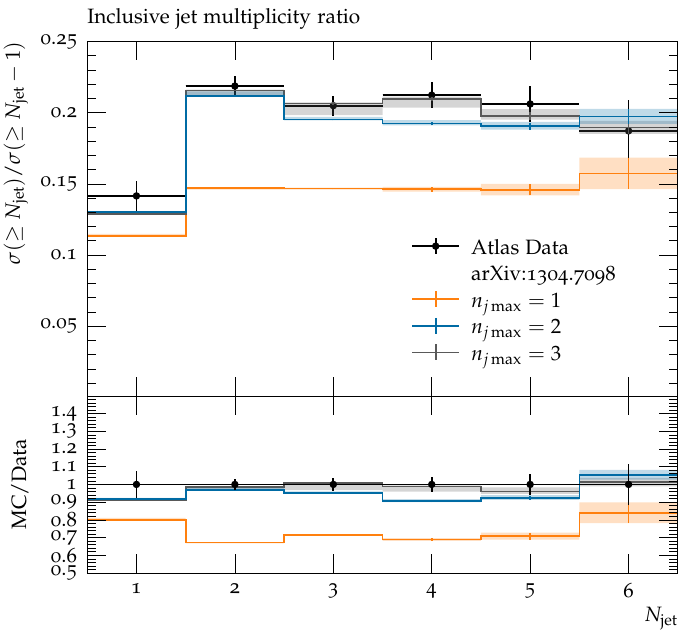}\\
  \caption{
    Multi-jet merged predictions from \Alaric in comparison to ATLAS measurements
    at 7 TeV~\cite{ATLAS:2013lly}, again in dependence on $n_{j\,\rm{max}}$. 
  See the main text for details.
    \label{fig:dy_jet_multi}}
\end{figure}
\begin{figure}[t]
  \centering
  \includegraphics[width=7.5cm]{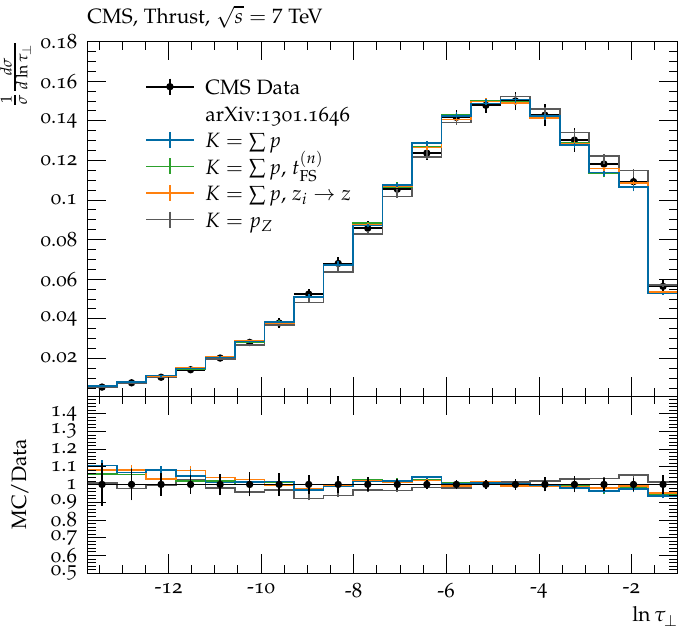}
  \includegraphics[width=7.5cm]{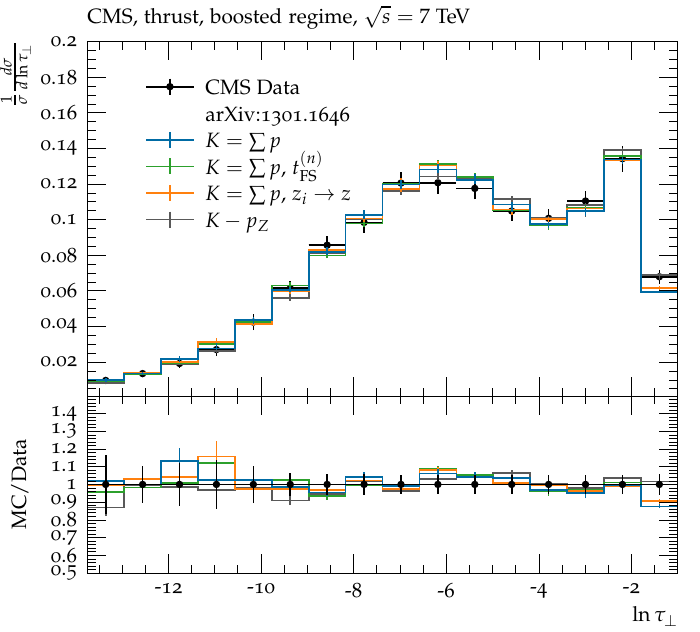}\\
  \includegraphics[width=7.5cm]{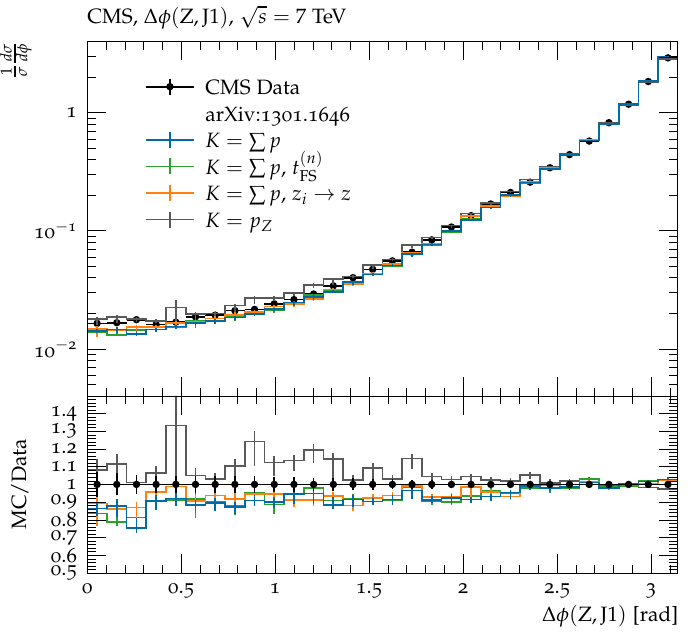}
  \includegraphics[width=7.5cm]{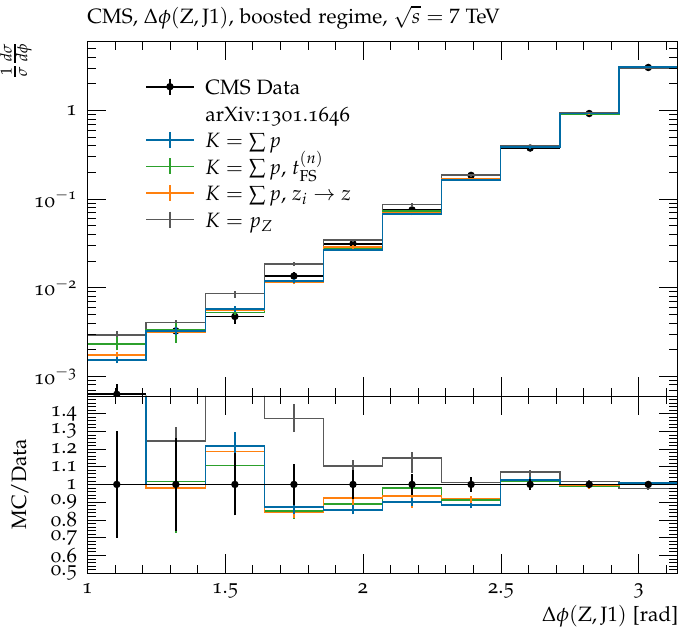}\\
  \caption{Multi-jet merged predictions from \Alaric in comparison to CMS
  measurements~\cite{CMS:2013lua}. See the main text for details.
    \label{fig:dy_shapes_correlations}}
\end{figure}
Figure~\ref{fig:dy_jet_multi} displays jet-multiplicity spectra in
$Z+$ jets multi-jet merged predictions from \Alaric in comparison to
measurements from the ATLAS collaboration~\cite{ATLAS:2013lly}.
There, electrons are required to have $|\eta|< 1.37$ or
$1.52<|\eta|<2.47$ and muons must be within $|\eta|<2.4$.  Both
electrons and muons must have $p_T>20$~GeV and are required to satisfy
the invariant mass constraint 66~GeV$\le m_{ll}\le$116~GeV.  The
leptons are dressed by photons within a cone of radius $R=0.1$ and
must satisfy $\Delta R_{ll}>0.2$.  Jets are reconstructed using the
anti-$k_T$ algorithm~\cite{Cacciari:2008gp} with a radius of $R=0.4$
and are required to have $|\eta|<4.4$ and $p_T>30$~GeV.  In addition,
they must be separated from leptons by $\Delta R>0.5$.  The left panel
shows the inclusive jet multiplicity distribution, and the right panel
displays the ratio of consecutive jet rates.  We present Monte-Carlo
results with increasing number of $n_{j\,\rm max}$ to exemplify that
the correct modeling of these distributions depends on the appropriate
coverage of the multi-jet phase space and the incorporation of the
tree-level matrix elements at sufficiently high final-state
multiplicity.  A computation with $n_{j\,\rm max}=1$ fails to describe
the experimental data, while the calculations with $n_{j\,\rm max}=2$
and $n_{j\,\rm max}=3$ are fairly similar.  In particular, the result
with $n_{j\,\rm max}=3$ is in good agreement with the jet multiplicity
ratio measurement above $N_\text{jet}=1$.  The uncertainty bands shown
in the figure represent the envelope of the statistical uncertainties
and the merging cut variations, with the merging cut varied between
5~GeV and 20~GeV.  We attribute the rate mismatch in the jet
multiplicity distribution above $N_\text{jet}=0$ and the corresponding
mismatch at $N_\text{jet}=1$ in the jet rate ratio to the missing
higher-order corrections, which are larger for the one-jet rate and
the subsequent jet rates than for the inclusive
process~\cite{Hoeche:2012yf,Neumann:2022lft}.

Figure~\ref{fig:dy_shapes_correlations} shows predictions from a 
$Z+$jets multi-jet merged computation with $n_{j\,\rm max}=2$ in
comparison with experimental measurements at 7 TeV c.m.-energy from
CMS~\cite{CMS:2013lua}.  The upper left panel shows the transverse
thrust distribution~\cite{Banfi:2010xy}, and the upper right panel
displays the same in the boosted region, where $p_{T,Z}>150$~GeV.  The
lower left panel shows the azimuthal decorrelation between the
Drell-Yan lepton pair and the leading jet, $\Delta\phi(Z,J1)$, and the
lower right panel shows the same in the boosted region, where
$p_{T,Z}>150$~GeV.  The Monte-Carlo predictions have been obtained
with the same settings as in Fig.~\ref{fig:dy_uncertainties}.  The
difference between the individual results is small, except for the
azimuthal decorrelation in the boosted regime, where the recoil
definition using only the di-lepton pair fails to describe the
small-$\Delta\phi$ region.  This is expected, because in the
boosted regime the Drell-Yan invariant mass does not provide the
largest scale in the process.

\subsection{Inclusive jet and di-jet production}
In this sub-section we compare results from a pure parton-shower
simulation, {\em without} applying any multi-jet merging, with \Alaric
against inclusive jet and dijet measurements from the ATLAS and CMS
collaborations.
\begin{figure}[t]
    \centering
    \includegraphics[width=.83\textwidth]{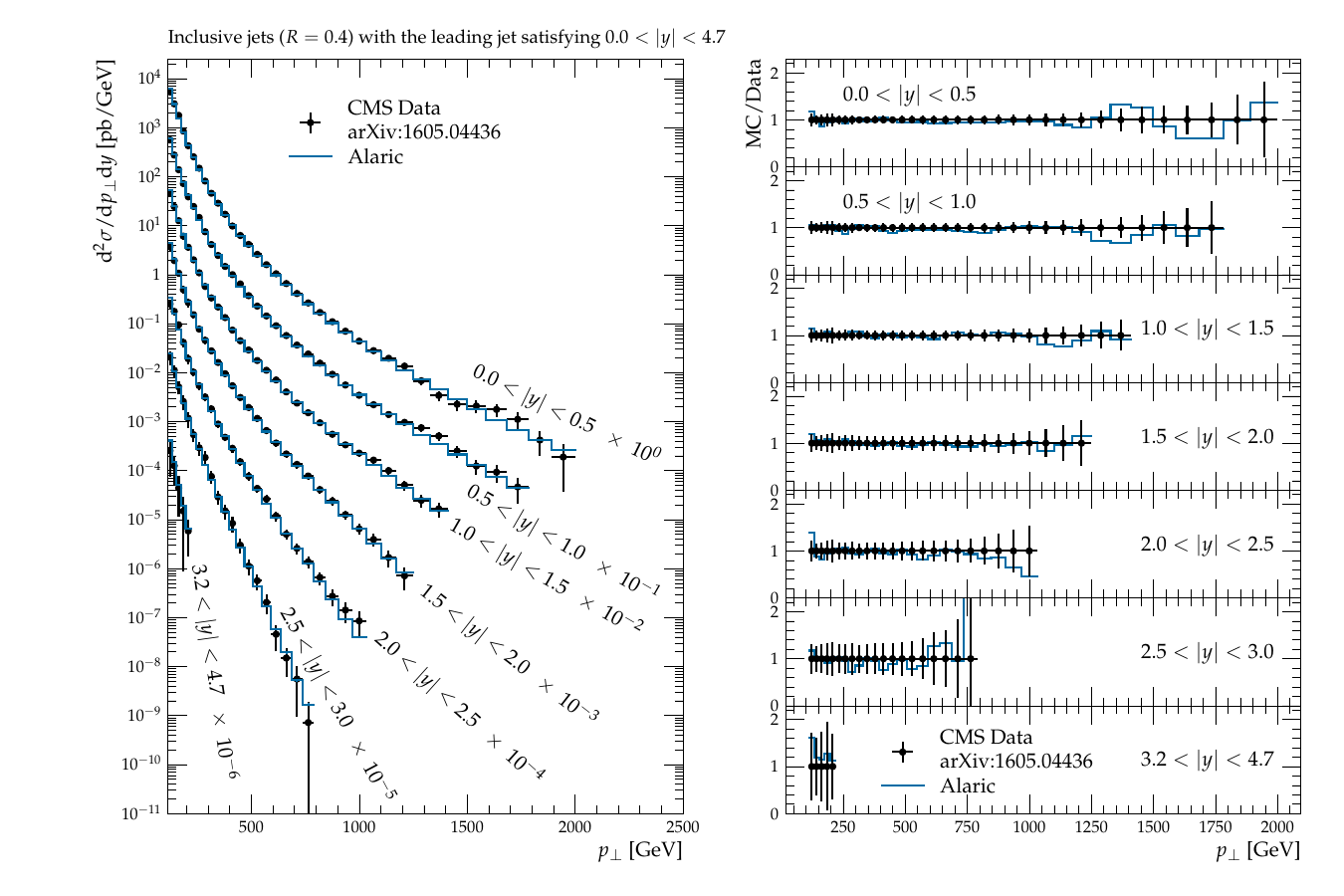}
    \caption{Transverse momentum spectrum of inclusive jets in different
      rapidity regions in proton-proton collisions at a center of mass energy of
      $13$~TeV. \Alaric predictions compared to data measured by CMS
      \cite{CMS:2016jip}. The left plot shows the full distributions while the
      panels on the right are the ratio to data.}
    \label{fig:incJetRates}
\end{figure}
The renormalization and factorization scales are chosen as $\mu_R =
\mu_F = H_T/4$, where $H_T$ denotes the scalar sum of the final state
transverse momenta.  The resummation scale ({\it i.e.}\ the parton
shower starting scale) is defined as $\mu_Q=p_\perp$, with $p_\perp$
the transverse momentum of the leading jet.  We compare to data measured
at the LHC at $\sqrt{s}=7~\text{TeV}$ and $\sqrt{s}=13~\text{TeV}$.
Hadronization corrections are included using the Lund model via an interface
to Pythia 8 \cite{Bierlich:2022pfr}.  We use the string fragmentation
parameters $a=0.4$, $b=0.36$ and $\sigma=0.3$.  To simulate the
underlying event we rely \Sherpa's default module
\cite{Gleisberg:2008ta}, based on the Sj\"ostrand–Zijl multiple-parton
interaction (MPI) model~\cite{Sjostrand:1987su}.  It is worth noting
that so far we have not produced a dedicated tune of hadronization
or underlying event parameters specifically for the \Alaric parton shower.

We start our discussion by firstly comparing, in
Fig.~\ref{fig:incJetRates}, \Alaric results to inclusive jet rates in
dependence on the transverse momentum of the leading jet, in several
bins of the leading jet rapidity.  The data were taken by the CMS
collaboration at $\sqrt{s}=13~\text{TeV}$~\cite{CMS:2016jip} and reach
energy scales up to $p_\perp\sim 2~\text{TeV}$ and rapidity values of
up to $|y| = 4.7$.  Our predictions are in good agreement with data,
which motivates us to investigate the details of the radiation pattern
in more detail.

\begin{figure}[t]
    \centering
    \includegraphics[width=.42\textwidth]{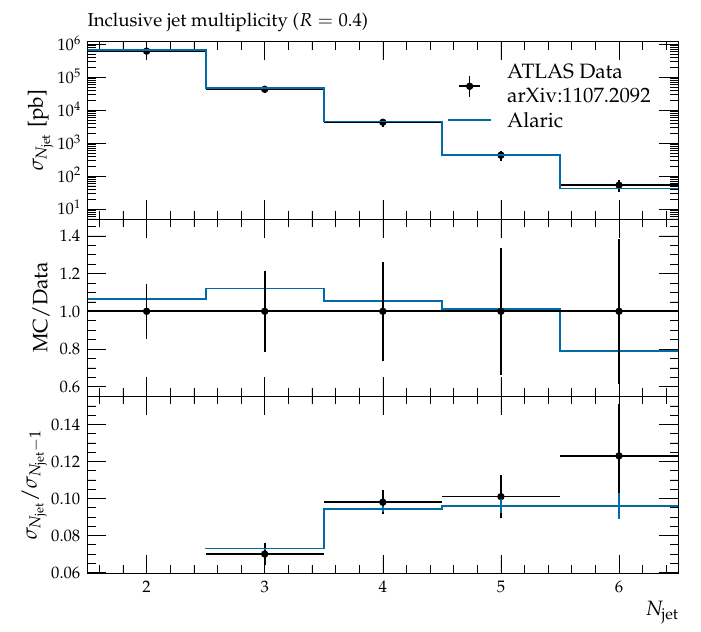}
    \includegraphics[width=.42\textwidth]{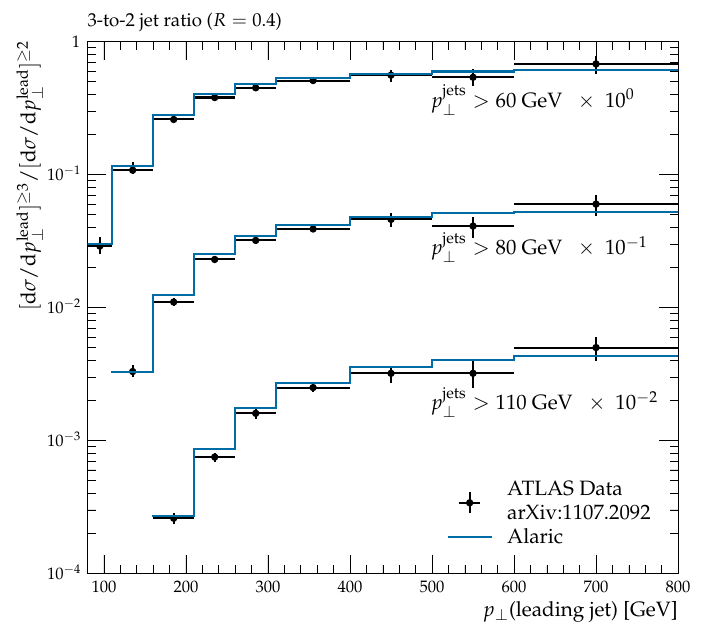}\\
    \includegraphics[width=0.42\textwidth]{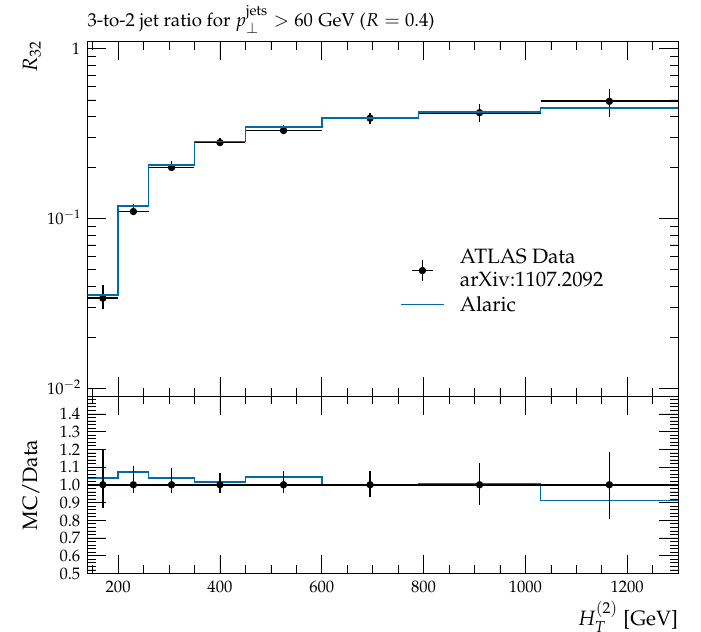}
    \includegraphics[width=0.42\textwidth]{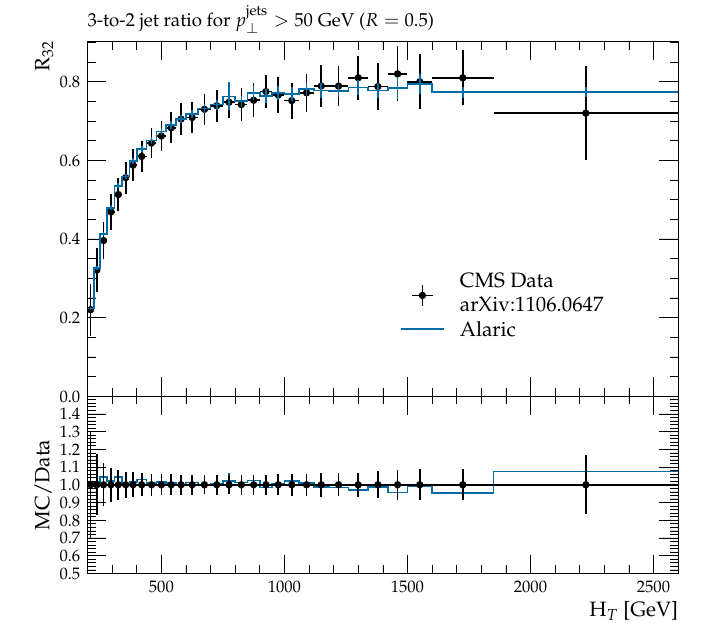}
    \caption{
    Inclusive jet multiplicity in inclusive jet production at
    $\sqrt{s}=7~\text{TeV}$ (upper left panel) with cross sections
    (top), the ratio of simulation and data (middle) and ratios
    between $N_\mathrm{jet}$ to $N_{\mathrm{jet}}-1$ jet rate
    (bottom).  The $N_\mathrm{jet}=3$ to $N_\mathrm{jet}=2$ rates
    (right) differential in the transverse momentum of the leading jet
    (upper right panel); both sets of data are taken
    from~\cite{ATLAS:2011qvj}.  Ratio of inclusive 3 jet over 2 jet
    rate $R_{32}$ at $\sqrt{s}=7~\text{TeV}$, as predicted by \Alaric
    and compared to measurements from ATLAS~\cite{ATLAS:2011qvj}
    (lower left) and CMS~\cite{CMS:2011maz} (lower right).}
    \label{fig:jetrates}
\end{figure}
We continue by comparing to the inclusive rates of jets produced in
the shower to data measured by ATLAS~\cite{ATLAS:2011qvj} at
$\sqrt{s}=7~\text{TeV}$.  The analysis constructs anti-$k_t$ jets with
a radius parameter of $R=0.4$, and requires at least one jet with a
transverse momentum of $p_\perp > 80~\text{GeV}$, while additional
jets are required to have $p_\perp > 60~\text{GeV}$.  All jets must
satisfy a rapidity requirement of $|y|<2.8$.  The comparison of the
cross sections for inclusive jet is presented in
Fig.~\ref{fig:jetrates}, starting from $N_\text{jet}=2$ and going up
to $N_\text{jet}=6$.  The \Alaric predictions slightly overestimate
the central value of the overall cross section for lower
multiplicities and tend to drop off somewhat faster for higher jet
rates than seen in data.  However, the predictions are consistent with
the data within the statistical uncertainties over the full range.
The ratio plot in the middle of the upper left panel of
Fig.~\ref{fig:jetrates} shows that the central value of the 3-jet rate
(although within the data uncertainty) is overestimated slightly more
than the inclusive 2-jet rate.  This effect is echoed in the bottom of
the upper left panel, where we plot the ratios of inclusive
$N_\text{jet}$ versus $N_\text{jet}-1$ rate.  In the upper right panel
of Fig.~\ref{fig:jetrates} we compare to data for the ratio of the 3-
to 2-jet rate, differential in the transverse momentum of the leading
jet, with different minimal requirements on the hardness of the
included jets.  We can see that the relative enhancement is mostly
constant over the full range of leading jet $p_\perp$.  A similar
dataset is available casting the 3-to-2-jet ratio as a function of the
scalar sum of the transverse momenta of the two leading jets,
$H_T^{(2)}$, or all jets, $H_T$.  We compare with 7 TeV data from
ATLAS \cite{ATLAS:2011qvj}, binned in $H_T^{(2)}$ in the lower left
panel of Fig.~\ref{fig:jetrates}, while the lower right plot compares
the shower with a similar measurement by the CMS
collaboration~\cite{CMS:2011maz} binned in $H_T$.  The CMS
measurement, likewise performed at $\sqrt{s}=7~\text{TeV}$, uses
anti-$k_t$ jets with an radius of $R=0.5$ and requires a transverse
momentum of at least $p_\perp^\text{jets} > 50~\text{GeV}$.  The
\Alaric predictions reproduce the data remarkably well, with
practically no discrepancy to either ATLAS or CMS data within the
uncertainty of the measurements.  This emphasizes that the \Alaric
algorithm can predict jet multiplicities and the 2-to-3 jet rate
with excellent quality from the parton shower alone.

\begin{figure}[t]
    \centering
    \includegraphics[width=.83\textwidth]{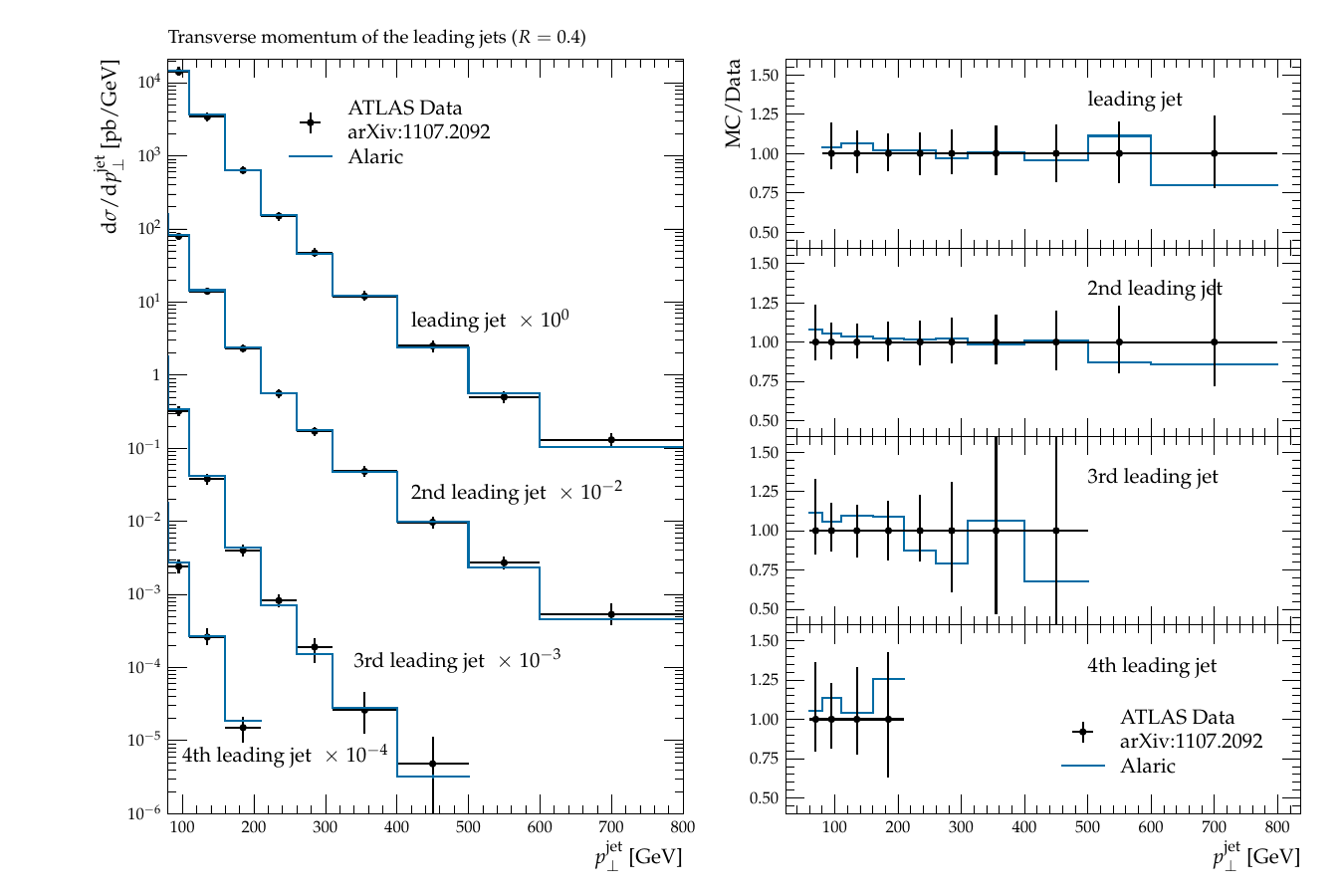}
    \includegraphics[width=.83\textwidth]{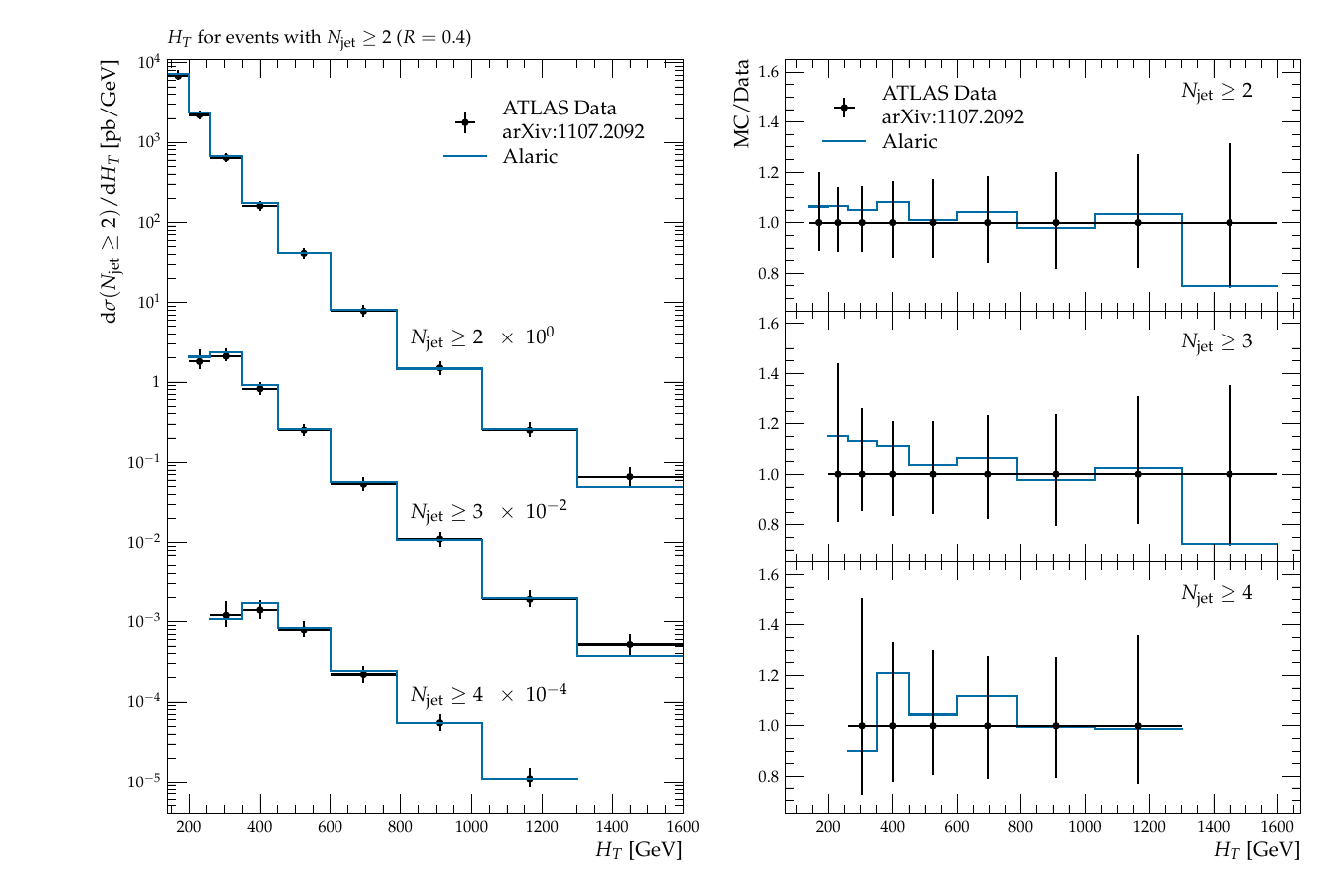}
    \caption{
    Transverse momentum spectra for the first 4 leading jets as
    predicted by \Alaric (upper left) and the ratio of simulation and
    data (upper right).  Cross sections differential in the $H_T$
    observable for events with at least 2, 3 and 4 hard jets (lower
    left) and the ratio of simulation and data (lower right), all with
    data from~\cite{ATLAS:2011qvj} taken at $\sqrt{s}=7~\text{TeV}$.}
    \label{fig:jetHT}
\end{figure}
We now turn to more differential measurements of jet properties.  The
upper panel of Fig.~\ref{fig:jetHT} shows the transverse momentum
spectra of the four leading jets (according to their $p_\perp$), as
predicted by \Alaric, and compares the results to 7 TeV measurements
from ATLAS~\cite{ATLAS:2011qvj}, providing data for transverse momenta
of the jets between $90~\text{GeV}$ and up to $800~\text{GeV}$ for the
leading and sub-leading jet(s).  The data are also available
differential in the $H_T$ observable, in the range
$180~\text{GeV}<H_T<1600~\text{GeV}$, separately for events containing
at least 2, 3 and 4 jets.  The comparison in the lower panel of
Fig.~\ref{fig:jetHT} presents a similar picture as the transverse
momentum data, the parton-shower result from \Alaric compares very
well over the entire range and for all considered multiplicities.
We again observe excellent agreement between our results and experimental
data, independent of the jet selection and over the full range of
transverse momentum studied.

\begin{figure}[t]
    \centering
    \includegraphics[width=.83\textwidth]{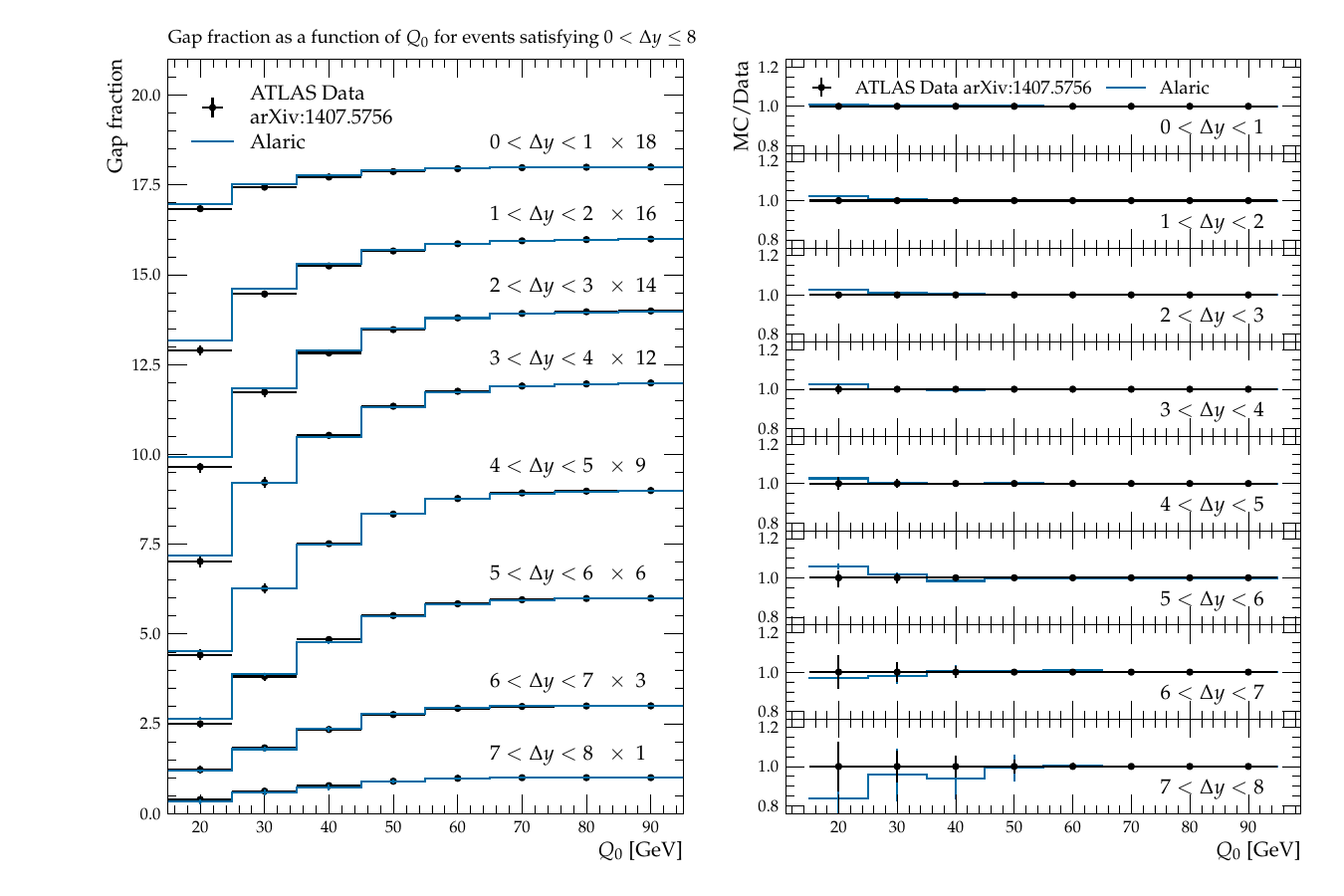}
    \caption{Gap fractions as a function of the cutoff $Q_0$ for different rapidity gaps $\Delta y$.
    \Alaric predictions are compared to data measured by ATLAS
    \cite{ATLAS:2014lzu} at $\sqrt{s}=7~\text{GeV}$. The left plot shows the
    full distributions while the panels on the right are the ratio to data.}
    \label{fig:gapFrac}
\end{figure}
While so far we have considered the transverse momenta and
multiplicity distributions of leading jets in the events, we next
analyze a class of observables sensitive to additional radiation in
the event.  To this end we consider non-global observables called gap
fractions, {\it i.e.}\ the fraction of events with no jets harder than
a cutoff $Q_0$ in the rapidity interval of size $\Delta y$ between the
two leading jets of a dijet system.  We compare our results to data
measured by the ATLAS experiment~\cite{ATLAS:2014lzu} at
$\sqrt{s}=7~\text{TeV}$ in Fig.~\ref{fig:gapFrac}.  This analysis uses
anti-$k_t$ jets with a radius of $R=0.6$, and the measurement is
presented in several $\Delta y$ bins starting from $0 \Delta y < 1$
ranging up to $7<\Delta y<8$.  We observe excellent agreement of the
data at larger $Q_0$ for the full range of $\Delta y$.  Only for the
smallest $Q_0$ values we find a slight excess of our parton-shower
predictions over the data.

\begin{figure}[t]
    \centering
    \includegraphics[width=0.42\textwidth]{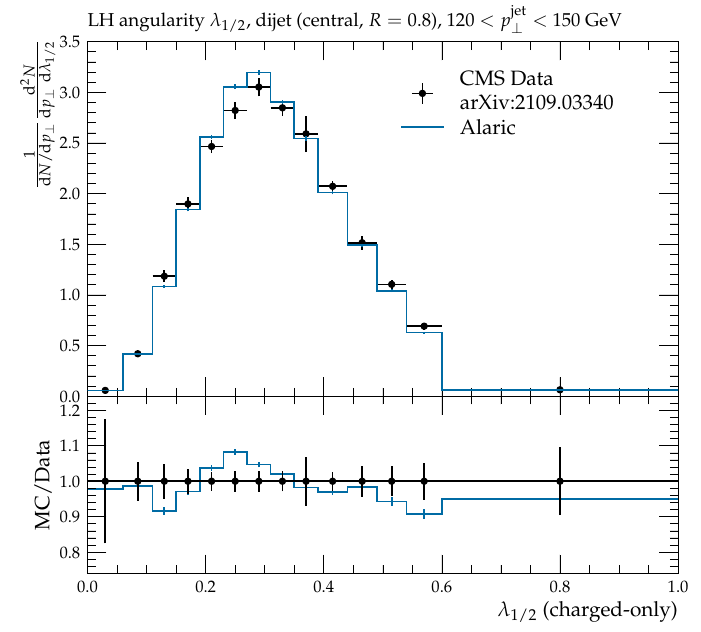}
    \includegraphics[width=0.42\textwidth]{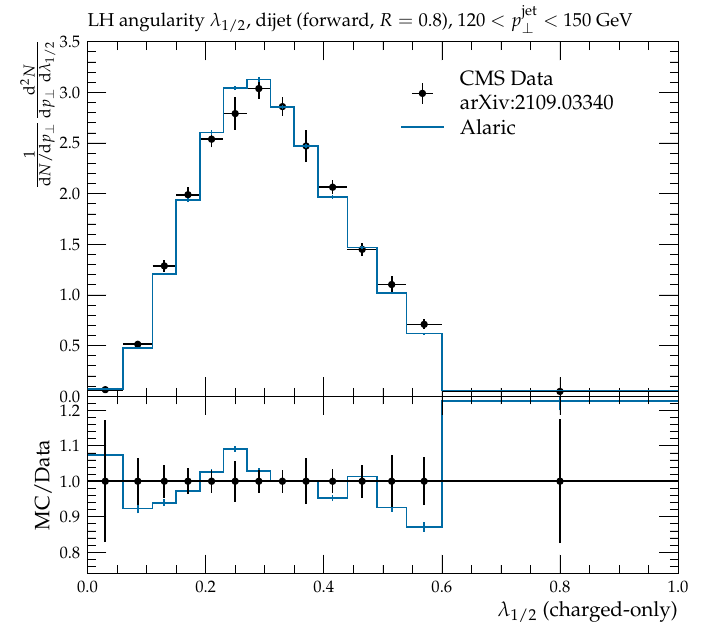}
    \caption{Les Houches angularity as measured by CMS \cite{CMS:2021iwu} at
      $\sqrt{s}=13~\text{TeV}$, on the more central (left) and more forward
      (right) of the two leading jets in dijet events.}
    \label{fig:lha}
\end{figure}
Finally, we highlight \Alaric's performance in describing the
intra-jet dynamics by presenting a comparison to a jet substructure
observable, in Fig.~\ref{fig:lha}.  The CMS collaboration has measured
several variants of angularities~\cite{CMS:2021iwu} in dijet events at
13~\text{TeV}.  This measurement has been studied extensively using
Sherpa in the past \cite{Caletti:2021oor, Reichelt:2021svh}.  For
brevity we restrict ourselves to showcasing the case of the so-called
Les Houches angularity \cite{Larkoski:2014pca, Andersen:2016qtm}
measured on charged particles in anti-$k_t$ jets with radius $R=0.8$.
We observe a similar level of agreement to the data as these earlier
studies, describing the general trend of the data but tentatively
producing somewhat narrower distributions than seen in data.
\FloatBarrier

\section{Conclusions}
\label{sec:outlook}
In this publication we introduced the novel \Alaric parton-shower for
simulating QCD radiation at hadron colliders, in particular the LHC.
We emphasized the importance of a correct identification of the momentum
fractions entering the purely collinear components of the splitting
functions.  We introduced a new evolution variable, which is defined
in the frame of the recoil momentum after the emission. This frame coincides
with the event frame in $e^+e^-\to$hadrons.  We also presented the first
multi-jet merging for the \Alaric parton shower.

We quantified the systematic uncertainties of the parton-shower predictions
due to various choices of recoil scheme, evolution and splitting parameters.
In a detailed comparison with experimental data from Drell-Yan lepton-pair 
production at the LHC we find that systematic uncertainties are relatively 
small. The only exception arises from the choice of recoil momentum, leading
to sizable uncertainties in some regions of phase space.  Driven by
the comparison with data, we argue that for generic LHC Drell-Yan plus
multi-jet events, the appropriate choice is a recoil vector that
includes all final-state particles.

We further highlighted the capabilities of the \Alaric algorithm
by comparing its predictions with an indicative range of relevant
observables, in particular the jet multiplicities in Drell-Yan lepton
pair production at the LHC. We also presented the first predictions
for inclusive jet and di-jet production.  We find that the quality
of the description of experimental data from the LHC is in line
with the current formal precision of the simulation.

In the near future, we will implement a next-to-leading order matching
procedure, and extend the leading-order merging to next-to-leading
order precision.  This will allow us to obtain state-of-the art
predictions for LHC measurements using the \Alaric algorithm.

\section*{Acknowledgments}
We thank Robert Szafron for discussions on the treatment of
sub-leading power corrections in SCET.
This research was supported by the Fermi National Accelerator
Laboratory (Fermilab), a U.S.\ Department of Energy, Office of
Science, HEP User Facility.  Fermilab is managed by Fermi Research
Alliance, LLC (FRA), acting under Contract No. DE--AC02--07CH11359.
This research used the Fermilab Wilson Institutional Cluster for code
development, testing, validation and production.  We are grateful to
James Simone for his support.  F.K.\ gratefully acknowledges funding
as Royal Society Wolfson Research fellow.  F.K.\ and D.R.\ are
supported by the STFC under grant agreement ST/P006744/1.

\bibliography{main}
\end{document}